\documentclass[11pt]{article}
\usepackage{verbatim,amsmath,amssymb}
\usepackage{epsfig,float,color}
\usepackage{epstopdf}
\usepackage{geometry}
\usepackage{setspace}
\usepackage{wrapfig}
\usepackage{hyperref}
\usepackage[utf8]{inputenc}
\usepackage{graphicx}
\usepackage{flushend}
\usepackage[linesnumbered,ruled]{algorithm2e}

\usepackage[font={footnotesize}]{caption}
\usepackage[font={footnotesize}]{subcaption}
\usepackage{footnote}
\usepackage{multicol}
\usepackage{multirow}
\usepackage{enumitem}   
\usepackage{soul}
\usepackage{framed}
\usepackage[titletoc,title]{appendix}
\usepackage{bm}
\usepackage{siunitx}

\usepackage[makeroom]{cancel} 
\usepackage{cite}
\usepackage{natbib}
\usepackage{hyperref}
\definecolor{darkblue}{rgb}{0,0,1}
\hypersetup{pdftex=true, colorlinks=true, breaklinks=true, linkcolor=magenta, menucolor=magenta, citecolor=magenta, urlcolor=magenta}

\usepackage[flushleft]{threeparttable}
\usepackage{tikz}
\usepackage{pgfplots}
\pgfplotsset{compat=newest}
\usetikzlibrary{positioning, arrows.meta}
\tikzset{%
	myarrow/.style = {-Stealth, shorten >=5pt}
}

\usetikzlibrary{calc, pgfplots.groupplots, backgrounds}
\usepgfplotslibrary{colorbrewer}
\usepackage{standalone}

\newcommand{\trr}[1]{{#1}^{\!\top}}
\newcommand{\pd}[2]{\frac{\partial #1}{\partial #2}}
\newcommand{\inv}[1]{{#1}^{\text{-}1}}

\newcommand{\mvect}[1]{\mathbf{#1}}
\newcommand{\mfield}[1]{\bm{#1}}
\usepackage[title]{appendix}
\usepackage{mathtools}
\newcommand{\verteq}{\rotatebox{90}{$\,=$}}
\newcommand{\equalto}[2]{\underset{\scriptstyle\overset{\mkern4mu\verteq}{#2}}{#1}}

\definecolor{darkblue}{rgb}{0,0,1}
\hypersetup{pdftex=true, colorlinks=true, breaklinks=true, linkcolor=darkblue, menucolor=darkblue, pagecolor=darkblue, citecolor=darkblue, urlcolor=darkblue}

\begin{document}
	
	\begin{center}
		\Large{\bf{Topological synthesis of fluidic pressure-actuated robust compliant mechanisms}}\\
		
	\end{center}
	
	\begin{center}
			\large{Prabhat Kumar\,$^{\star,\,\ddagger,}$\footnote{Corresponding author, email: \url{pkumar@mae.iith.ac.in}} and Matthijs Langelaar\,$^{\dagger}$} \\
		\vspace{4mm}
		
		\small{\textit{$\star$Department of Mechanical and Aerospace Engineering, Indian Institute of Technology Hyderabad, 502285, India}}\\
			\vspace{2mm}	
		\small{\textit{$\ddagger$ Department of Mechanical Engineering, Indian Institute of Science, Bengaluru 560012, Karantaka, India}}\\
				\vspace{2mm}
		\small{\textit{$\dagger$Department of Precision and Microsystems Engineering, Delft University of Technology,2628CD Delft, The Netherlands}}\\

		\vspace{4mm}
Published\footnote{This pdf is the personal version of an article whose final publication is available at \href{https://www.sciencedirect.com/science/article/pii/S0094114X22001367}{Mechanism and Machine Theory}}\,\,\,in \textit{Mechanism and Machine Theory}, 
\href{https://doi.org/10.1016/j.mechmachtheory.2022.104871}{DOI:10.1016/j.mechmachtheory.2022.104871} \\
Submitted on 12~February 2022, Revised on 14~March 2022, Accepted on 30~March 2022
		
	\end{center}
	
	\vspace{3mm}
	\rule{\linewidth}{.15mm}
	{\bf Abstract:}
	This paper presents a robust density-based topology optimization approach for synthesizing pressure-actuated  compliant mechanisms. To ensure functionality under manufacturing inaccuracies, the robust or three-field formulation is employed, involving dilated, intermediate and eroded realizations of the design.  Darcy's law  in conjunction with a conceptualized drainage term is used to model the pressure load as a function of the design vector. The consistent nodal loads are evaluated from the obtained pressure field using the standard finite element method.  The objective and load sensitivities are obtained using the adjoint-variable approach. A multi-criteria objective involving both the stiffness and flexibility of the mechanism is employed in the robust formulation, and min-max optimization problems are solved to obtain pressure-actuated inverter, gripper, and contractor compliant mechanisms with different minimum feature sizes. Limitations of the linear elasticity assumptions while designing mechanisms are identified with high pressure loads. Challenges involved in designing finite deformable pressure-actuated compliant mechanisms are presented. \\
	

	\vspace{-4mm}
	\rule{\linewidth}{.15mm}
	
\section{Introduction}

Compliant mechanisms (CMs) are established concepts in industry and academia offering various advantages over  traditional linkage-based counterparts, e.g., less wear and tear, low manufacturing and assembly cost, repeatability and high precision, lack of frictional losses, to name a few. Due to such promising advantages, their usage is continuously rising in a wide variety of applications \citep{frecker1997topological,howell2001compliant,kumar2019compliant,kumar2021topology,ananthasuresh2021art}. These mechanisms characterized via monolithic designs are termed CMs, since their functionality arises from the elastic deformations of their flexible (compliant) members in response to the input forces. Finding the optimum balance between output deformation and stiffness when designing CMs is a nontrivial task \citep{cao2018topology}. Topology optimization (TO) has been shown to be an effective approach for designing such mechanisms \citep{zhu2020design}. TO is a computational design technique able to achieve the optimized material distribution within a given design domain by extremizing the conceptualized (desired) objectives under a known set of physical and geometrical constraints \citep{sigmund2013topology}. In a general structural setting, the design domain is parameterized using finite elements (FEs). Each FE is assigned a material density design variable $\rho_i\in[0,\,1]$.  Here, $\rho_i=1$ and  $\rho_i=0$ represent solid and void states of the $i^\text{th}$ FE, respectively. Ideally,  FEs with $\rho=1$ should constitute the optimized CMs. 

Actuating forces of CMs can be recognized as either \textit{design-dependent}, e.g., pneumatic, hydraulic pressure loads, or  \textit{design-independent}, e.g., constant forces.  Design-dependent pressure loads\footnote{We henceforth for brevity write pressure loads instead of design-dependent pressure loads.} alter their magnitude, location and/or direction as the design boundary on which they act evolves during the TO process. Consequently, such loads pose many challenges for the TO formulation, e.g., locating a valid surface to apply the loads, relating pressure field to the design vector, and evaluating consistent nodal forces and their sensitivities with respect to the design vector \citep{kumar2020topology}.

\begin{figure*}
	\centering
	\includegraphics[scale=1]{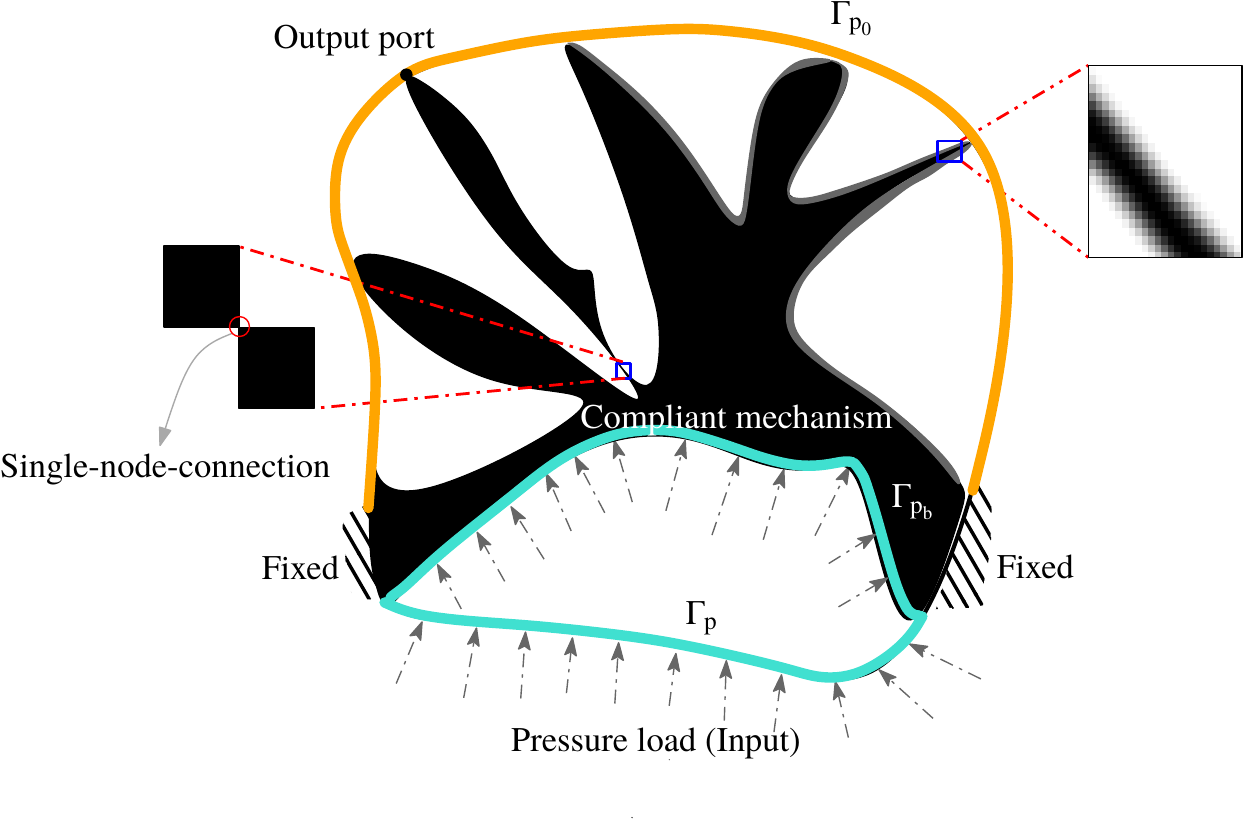}
	\caption{Schematic diagram of a pressure-actuated compliant mechanism. $\Gamma_{\text{p}_{0}}$ and $\Gamma_\text{p}$ are the boundaries with zero and finite pressure loading, respectively. $\Gamma_{\text{p}_{b}}$ is the final pressure boundary for the optimized CM. A single-node hinge and a region with gray FEs are also depicted.} \label{fig:Schematic}
\end{figure*}

\textit{Pressure-actuated CMs (Pa-CMs)} constitute a relatively novel category of mechanisms that find application in e.g. pneumatically or hydraulically actuated soft robots (mechanisms) \citep{chen2018topology,luo2020topology}. Note however that, in order to attain maximum flexibility, CMs designed via classical TO are prone to exhibit single-node-connected hinges and gray density FEs (0$<\rho<$1) in the optimized designs \citep{sigmund1997design,yin2003design}. Such features cannot be realized, and in order to render the design manufacturable, post-processing of the optimized geometry is necessary which can severely affect the optimized performance. This challenge is found in Pa-CMs designed by TO as well, where it can be even more detrimental given the close relation between boundary shape and loading. A schematic figure of a Pa-CM is depicted in Fig.~\ref{fig:Schematic} wherein the pressure loading boundary moves from its initial surface $\Gamma_{\text{p}}$ to the final (optimized) surface $\Gamma_{\text{p}_{b}}$. Furthermore, Fig.~\ref{fig:Schematic} depicts a single-node hinge and region with gray FEs. Given the negative effect such features have on the manufacturability and performance of Pa-CMs, it is important to control and avoid them. The need to generate Pa-CM designs whose actual performance closely matches the simulated optimized performance forms the motivation for the present study.

\section{Background and Approach}
\citet{Hammer2000} were first to present an approach to design pressure-loaded structures. A fictitious thermal loading setting was exploited to solve pressure-loaded design problems by \citet{Chen2001}.  \citet{Sigmund2007} used the mixed-finite element method \cite{zienkiewicz2005finite} with a three-phase material (solid, void, fluid) formulation in their approach. The mixed-finite element approaches require satisfaction of the Babuska-Brezzi condition for the stability in the FE analysis \cite{zienkiewicz2005finite}. \citet{chen2001advances} employed the approach presented in \citet{Chen2001} to design Pa-CMs. \citet{Panganiban2010} used a nonconforming FE method which is not a standard FE method.  \citet{vasista2012design} employed the solid isotropic material with penalization (SIMP) and the moving isosurface threshold methods in their approach.  \citet{de2020topology}  also employed the method proposed by \citet{Sigmund2007}. In our previous study \cite{kumar2020topology}, the authors presented a design method using Darcy's law in conjunction with a  drainage term. In that work we also demonstrated the importance of  load sensitivities in designing of the Pa-CMs.  The  method proposed in \cite{kumar2020topology} uses the standard FE formulation, provides consistent sensitivities and was found to work well in generating Pa-CMs for two- and three-dimensional as well \citep{kumar2020topology3Dpressure}. Therefore, it will also be used in the present study. 

\begin{figure*}
	\begin{subfigure}[t]{0.48\textwidth}
		\centering
		\includegraphics[scale=0.5]{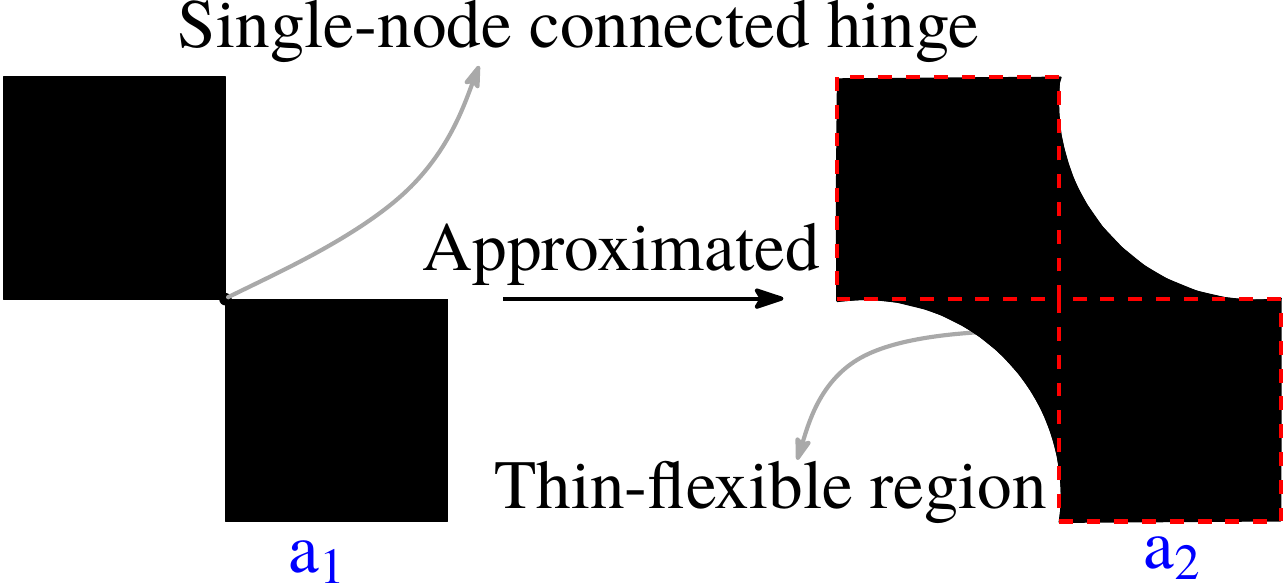}
		\caption{}
		\label{fig:pointconnection}
	\end{subfigure}
	\begin{subfigure}[t]{0.48\textwidth}
		\centering
		\includegraphics[scale=0.50]{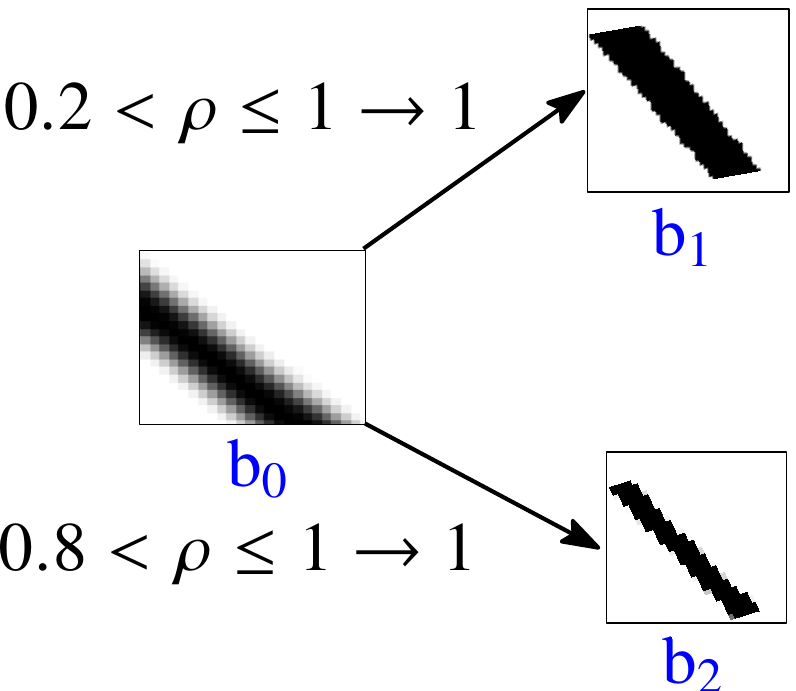}
		\caption{}
		\label{fig:thresolding}
	\end{subfigure}
	\caption{ Schematic diagrams for a single-node connection and design thresholding are depicted in (\subref{fig:pointconnection}) and (\subref{fig:thresolding}), respectively. (\subref{fig:pointconnection}) shows one of the ways to approximate the design around a single-node connection into a thin-flexible region and therefore, the fabricated mechanisms may  have lower  performance than the  corresponding numerical results. (\subref{fig:thresolding}) depicts how the different threshold material densities  result in different material layouts for manufacturing and thus, performance of the fabricated Pa-CM depends upon the threshold density one chooses during design extraction.} \label{fig:Schematicdiagram}
\end{figure*}

For the design optimization process to be useful and reliable, it is important that as-fabricated Pa-CMs perform similar to the prediction made by the numerical simulation used in the optimization process. However, with existing methods  a significant decline in actual performance can arise compared to the numerical predictions. This is primarily due to three factors: (A) inaccurate and/or approximate conversion of one-node-connected hinges to thin-flexible regions (Fig.~\ref{fig:pointconnection}) and inaccuracies introduced by unrealistic representation of thin, flexible regions in FE models used in TO, (B) CMs being overly sensitive to manufacturing inaccuracies or arbitrariness in design extraction (Fig.~\ref{fig:thresolding}), and (C) the use of small displacement analysis and linear elasticity assumptions for the Pa-CM designs. 

Firstly, one-node-connected hinges (Fig.~\ref{fig:pointconnection}), artificially stiff locations, appear due to deficiencies in the FE analysis with quadrilateral FEs that permit load transfer with zero rotational stiffness \citep{sigmund1997design,yin2003design}. Such hinges pose challenges in  accurate design interpretation of the optimized mechanisms, since real compliant hinges will always have a finite rotational stiffness \citep{sigmund1997design,yin2003design}.  One of the methods for approximating a one-node-connected location for fabrication may be as depicted in Fig.~\ref{fig:pointconnection}, which results in a thin-flexible region and thus, the performance of the numerical design (Fig.~\ref{fig:pointconnection}-a$_1$) will differ from that of the fabricated one (Fig.~\ref{fig:pointconnection}-a$_2$). Various ways have been proposed to prevent formation of single-node-connected hinges in CMs \citep{poulsen2003new,yin2003design,saxena2007honeycomb,wang2011projection,cao2015hybrid}, but these have not yet been applied to and evaluated for Pa-CM TO.

Secondly, in a standard density-based  TO setting with a gradient-based optimizer, it is difficult to obtain pure 0-1 solutions (Fig.~\ref{fig:Schematic}). Therefore, extraction of the optimized designs based on the considered density threshold (Fig.~\ref{fig:thresolding}) is required, which invariably alters the final designs and thus, the performances with respect to the numerical predictions. Fig.~\ref{fig:thresolding} illustrates a scenario to indicate how the different threshold material densities lead to different material (contour) layouts (Fig.~\ref{fig:thresolding}-b$_1$ and Fig.~\ref{fig:thresolding}-b$_2$) and thus, the corresponding fabricated Pa-CM and potentially also its loading (significantly) differs from that obtained via TO. 

Thirdly, a typical Pa-CM or CM may experience large deflection and also, contact between branches, i.e., self-contact \citep{kumar2017implementation} and external/mutual contact \citep{kumar2016synthesis,kumar2019computational} thus, TO design approaches must include nonlinear mechanics (with contact formulation) to predict mechanism performance \citep{kumar2016synthesis,kumar2019computational}. However, nonlinear structural analysis poses various challenges in TO \citep{van2014element,kumar2021topology}, which can even get more pronounced in combination with pressure loads whose magnitude, direction and/or location vary and follow the surfaces/facets where they are applied. To model the characteristics of  pressure loads, one needs to include the follower force concepts in the design approach \citep{zienkiewicz2005finite}, which demands a dedicated and in-depth investigation within a TO setting, which is out of the scope of this paper. Therefore, instead of addressing this point at the TO stage, we choose to investigate and assess its influence based on the Pa-CM designs generated using linear modeling.

To address the outlined challenges, in this study we take the following approaches:
\begin{enumerate}
	\item The robust or three-field formulation \citep{wang2011projection} is adopted to address problems A~(approximate conversion of one-node-connected hinges to thin-flexible regions) and B~(CMs being overly sensitive to manufacturing inaccuracies or arbitrariness in design extraction), by its ability to impose a minimum length scale and to reduce the sensitivity of the final design to post-processing or manufacturing errors.
	\item  Nonlinear FEA  is used to analyze optimized Pa-CM designs with a neo-Hookean hyperelastic material model and increased pressure loads to investigate the large deformation behavior and to determine the limitations of the linear elastic assumptions. 
\end{enumerate}
While the robust formulation \citep{wang2011projection} is expected to solve factors A and B given results reported in literature \citep{wang2011projection}, this has as of yet not been investigated or confirmed for Pa-CM, which is the specific interest of this study. Next to this, the nonlinear FEA i.e. geometric nonlinearity (due to finite deformation in the mechanisms) and material nonlinearity (as rubber-like materials are typically used to fabricate such mechanisms, cf.  \citep{schmitt2018soft}) study  is not intended as a solution to the lack of large-displacement analysis during Pa-CM TO, but our aim is to provide a quantitative assessment of the severity of the error introduced by this simplification, in representative design cases and also, the challenges involved in designing finite-deformable Pa-CMs. In addition, nonlinear finite element analyses are used to assess the sensitivity of standard and robust Pa-CMs to design extraction choices/ manufacturing errors. Due to computational limitations and for clarity of presentation this study has been carried out in a 2D setting. Its findings nonetheless are expected to apply to general 3D cases as well.

In particular, the current paper offers following new aspects:
\begin{itemize}
	\item  A robust topology optimization approach to optimize design-dependent fluidic Pa-CMs. The Darcy law in conjunction with the conceptualized drainage term \citep{kumar2020topology} is employed for the pressure field modeling, whereas material layout modeling is performed using the three-field (dilated, intermediate and eroded fields) formulation \citep{wang2011projection}.
	\item Demonstration of the robustness and efficacy of the proposed method  by designing various pressure-driven compliant mechanisms, e.g., inverter, gripper, and contractor mechanisms.  The optimized compliant mechanisms contain a large space for pressure to inflate them like soft robots, which makes the optimized mechanisms uniquely different from the other pressure-actuated compliant mechanisms previously designed by topology optimization.
	\item Assessment of the optimized Pa-CMs obtained via linear elastic assumptions under high pressure loads. Several challenges are demonstrated and discussed for designing finite deformation fluidic Pa-CMs. 
	\item A method to extract the optimized designs to facilitate CAD modeling for further analyses and 3D printing/fabrication.
\end{itemize}

This paper is organized  as follows. Sec.~\ref{Sec:Desingdependentpressureloads} presents modeling of the pressure loads as a function of the design vector using the Darcy law with a drainage term, in line with \citet{kumar2020topology}. Using a transformation matrix, the consistent nodal loads are evaluated. The TO formulation with the robust approach together with the corresponding sensitivity analysis is described in Sec.~\ref{Sec:TOformulation}. Next, in Sec.~\ref{Sec:Numericalexamples} numerical examples of designing robust pressure-actuated inverter and gripper mechanisms are presented. The optimized Pa-CM designs are extracted, and nonlinear FE analyses are performed in ABAQUS with high pressure loads to investigate large deformation behavior of the CMs. Lastly, conclusions are drawn in Sec.~\ref{Sec:Closure}. 

\section{Design-dependent pressure load modeling}\label{Sec:Desingdependentpressureloads}

In this section, modeling of the pressure field as a function of the design variables, finite element formulation and consistent nodal loads evaluation are summarized for the sake of self-consistency. For a detailed description, we refer to our previous paper \citep{kumar2020topology}.

The material boundaries of a given design problem evolve as TO progresses. Thus, it becomes challenging especially at the beginning of the optimization to locate an appropriate boundary to apply the fluidic pressure load. In addition, a design-dependent and continuous pressure field are expected to  help the TO process.  Further, at the initial stage of the optimization, one can consider each element as a porous medium, and  boundaries with the prescribed input pressure and zero pressure loads are  already provided. Therefore, the Darcy law is adopted herein to model the pressure field wherein the flow coefficient of each element is interpolated using a smooth Heaviside function \citep{kumar2020topology,kumar2020topology3Dpressure}. As per the Darcy law, one evaluates flux $\bm{q}$ in terms of the pressure gradient $\nabla p$, the permeability $\kappa$ of the medium and the fluid viscosity $\mu$ as
\begin{equation}\label{eq:Darcyflux}
	\bm{q} = -\frac{\kappa}{\mu}\nabla p = -K(\bar{\rho}) \nabla p,
\end{equation}
where $\bar{\rho}$ and $K(\bar{\rho})$ represent the physical density (see Sec.~\ref{Sec:TOformulation}) and  the flow coefficient of an FE, respectively. In a typical density-based TO setting, an FE displays two states, therefore the actual flow coefficient $K(\bar{\rho_e})$ of an FE is determined using the flow coefficients associated to its solid and void phases interpolated by a smooth Heaviside projection function $\mathcal{H}(\bar{{\rho_e}},\,\beta_\kappa,\,\eta_\kappa)$ as
\begin{equation}\label{Eq:Flowcoefficient}
	K(\bar{\rho_e}) = K_v\left(1-(1-\epsilon) \mathcal{H}(\bar{{\rho_e}},\,\beta_\kappa,\,\eta_\kappa)\right),
\end{equation}
where $\mathcal{H}(\bar{{\rho_e}},\,\beta_\kappa,\,\eta_\kappa) = \frac{\tanh{\left(\beta_\kappa\eta_\kappa\right)}+\tanh{\left(\beta_\kappa(\bar{\rho_e} - \eta_\kappa)\right)}}{\tanh{\left(\beta_\kappa \eta_\kappa\right)}+\tanh{\left(\beta_\kappa(1 - \eta_\kappa)\right)}}$, and $\epsilon=\frac{K_s}{K_v}$ is the flow contrast \citep{kumar2020topology3Dpressure}. $K_s$ and $K_v$ indicate the flow coefficients of the solid and void states, respectively. Further, $\eta_\kappa$ and $\beta_\kappa$ control the step position and the slope of $K(\bar{\rho_e})$, respectively.  In addition, a drainage term $Q_\text{drain}$ conceptualized in \citet{kumar2020topology} and numerically qualified  in \citet{kumar2020topology3Dpressure} is employed that helps achieve a localized pressure gradient at solid-void interfaces. It is defined in terms of a drainage coefficient $D(\bar{\rho_e})$, instantaneous pressure field $p$ and output pressure $p_{\text{ext}}$ as 
\begin{equation}
	{Q}_\text{drain} = -D(\bar{\rho_e}) (p - p_{\text{ext}}),
\end{equation}
where the drainage coefficient $D(\bar{\rho_e}) =  D_{\text{s}}\mathcal{H}(\bar{{\rho_e}},\,\beta_d,\,\eta_d)$.  $\beta_\text{d}$ and $\eta_\text{d}$ are two parameters that control the values of $D(\bar{\rho_e})$. $D_\text{s}$ is the drainage coefficient of a solid FE, which is equal to \citep{kumar2020topology}
\begin{equation}\label{Eq:hsrelation}
	D_\text{s} =\left(\frac{\ln{r}}{\Delta s}\right)^2 K_\text{s}, 
\end{equation} 
where $r$ is the ratio of input pressure at depth $\Delta$s, i.e., $p|_{\Delta s} = rp_\text{in}$ and the penetration depth $\Delta s$ can be set equal to the width or height of a few FEs. Using $Q_\text{drain}$, Eq.~\ref{eq:Darcyflux} transpires  per \citet{kumar2020topology} as 
\begin{equation}\label{Eq:stateequation}
	\nabla\cdot\bm{q} -Q_\text{drain} = 0.
\end{equation} 
In a discrete FE setting, one writes the weak form of Eq.~\ref{Eq:stateequation} for an FE with domain $\Omega_e$ as \citep{kumar2020topology}
\begin{equation} \label{Eq:PDEsolutionpressure}
	\begin{aligned}
		\equalto{\underbrace{\int_{\Omega_e}\left( K~ \trr{\mvect{B}}_p \mvect{B}_p   + D ~\trr{\mvect{N}_p} \mvect{N}_p \right)d {\Omega_e}}_{\mvect{A}_e}~\mvect{p}_e} 
		{\overbrace{\int_{\Omega_e}~D~\trr{\mvect{N}}_p p_\text{ext} ~~d {\Omega_e} -
				\int_{\Gamma_e}~ \trr{\mvect{N}}_p \mvect{q}_\Gamma \cdot \mvect{n}_e~~d {\Gamma_e}}^{\mvect{f}_e}}
	\end{aligned}
\end{equation}
where $\mvect{p}_e$ is the pressure field to be evaluated, $\mvect{B}_p =\nabla\mvect{N}_p$ with $\mathbf{N}_\text{p} = [N_1,\, N_2,\,N_3,\,N_4]$ are the bi-linear shape functions for a quadrilateral FE. $\mvect{Ap} = \mvect{f}$ is the global form of Eq.~\ref{Eq:PDEsolutionpressure} with $\mvect{p}$ is the global pressure load vector. In this work, $p_\text{ext}$ and $\mvect{q}_\Gamma$ are set to zero, therefore $\mvect{f}=\mvect{0}$, i.e., $\mvect{Ap}=\mvect{0}$. Using the obtained global pressure field $\mvect{p}$, the consistent global nodal forces $\mvect{F} = -\mvect{T}\mvect{p}$ are determined using a transformation matrix $\mvect{T}$ whose elemental form $\mvect{T}_e$ is related to that of nodal force  $\mvect{F}_e$ as  \citep{kumar2020topology}
\begin{equation}\label{Eq:Forcepressureconversion}
	\mvect{F}^e = \mvect{T}_e\,\mvect{p}_e = -\int_{\mathrm{\Omega}_e} \trr{\mvect{N}}_\mvect{u} \mvect{B}_p  d {\mathrm{\Omega}_e}\, \mvect{p}_e,
\end{equation}
where $\mathbf{N}_\mathbf{u} = [N_1\mathbf{I},\, N_2\mathbf{I},\,N_3\mathbf{I},\,N_4\mathbf{I}]$ with $\mathbf{I}$  the identity matrix in $\mathcal{R}^2$. In summary, one employs Eq.~\ref{Eq:PDEsolutionpressure} for determining the pressure field, whereas the corresponding consistent nodal force vector for an FE is evaluated using Eq.~\ref{Eq:Forcepressureconversion}. Note that through the use of smooth Heaviside functions the loads are a differentiable function of the density variables. This allows performing load sensitivity analysis readily, as detailed in \citet{kumar2020topology} and further elaborated in Sec.~\ref{Subsec:Sensitivityanalysis}.
	
	\section{Topology optimization formulation}\label{Sec:TOformulation}
	
	The three-field ($\bm{\rho},\,\tilde{\bm{{\rho}}},\,\bar{\bm{\rho}}$) representation of the design domain is considered \citep{lazarov2016length}. The filtered design variable $\tilde{\rho_j}$ of element~$j$ is determined using weighted average of the design variables $\rho$ pertaining to neighboring FEs lying within a circular region of radius $r_\text{fill}$ \citep{bruns2001topology}. Mathematically, 
	\begin{equation}\label{Eq:densityfilter}
		\tilde{\rho_j} = \frac{\sum_{k=1}^{N_e} v_k \rho_k w(\mvect{x}_k)}{\sum_{k=1}^{N_e} v_k w(\mvect{x}_k)} 
	\end{equation}
	where $N_e$ is the total number of neighboring elements of the $j^\text{th}$ FE, and $v_k$ is the volume of neighboring element~$k$. The weight function $w(\mvect{x}_k)= \max\left(0,\,1-\frac{d}{r_\text{fill}}\right)$, wherein $d = ||\mvect{x}_j -\mvect{x}_k||$ is a Euclidean  distance between centroids $\mvect{x}_j$ and $\mvect{x}_k$ of elements  $j$ and $k$, respectively. $r_\text{fill}$ is called  filter radius for the considered design problems. The derivative of $\tilde{\rho_j}$ (Eq.~\ref{Eq:densityfilter}) with respect to $\rho_k$  is
	\begin{equation}\label{Eq:derivativefilteractual}
		\pd{\tilde{\rho_j}}{\rho_k} = \frac{v_k w(\mvect{x}_k)}{\sum_{i=1}^{N_e}v_i w(\mvect{x}_i)}. 
	\end{equation}
	The physical  design variable $\bar{\rho}_j$  is defined as \citep{wang2011projection}
	\begin{equation}\label{Eq:projectionfilter}
		\bar{\rho_j}(\tilde{\rho_j},\,\beta,\,\eta)  = \frac{\tanh{\left(\beta \eta\right)} + \tanh{\left(\beta (\tilde{\rho_j}-\eta)\right)}}{\tanh{\left(\beta \eta\right)} + \tanh{\left(\beta (1-\eta)\right)}},
	\end{equation}

	where $\beta\in[0,\,\infty)$ and $\eta\in[0,\,1]$ control the steepness and the threshold of the projection function, respectively. To achieve the optimized solutions close to black and white designs,  typically $\beta$ is increased from an initial value~$\beta_\text{int}=1$ to a maximum value $\beta_{\max}$ using a continuation strategy. $\eta=0$  ensures the minimum length scale on the solid phase \citep{wang2011projection}, whereas that of solid phase is obtained using $\eta=1$. Note that when using $\eta=0$ and $\eta= 1$, Eq.~\ref{Eq:projectionfilter} yields the Heaviside step approximation function  given in \citet{guest2004achieving} and the modified Heaviside step approximation function mentioned in \citet{sigmund2007morphology}, respectively. The derivative of $\bar{\rho_j}$ with respect to $\tilde{\rho_j}$ is 
	\begin{equation}\label{Eq:derivativeprojectionfilter}
		\pd{\bar{\rho_j}}{\tilde{\rho_j}} = \beta\frac{1-\tanh(\beta(\tilde{\rho_j} -\eta))^2}{\tanh{\left(\beta \eta\right)} + \tanh{\left(\beta (1-\eta)\right)}}.
	\end{equation}
	Having noted the derivatives in Eqs.~\ref{Eq:derivativefilteractual} and \ref{Eq:derivativeprojectionfilter}, the chain rule is used to determine the derivatives of a function $f$ with respect to $\rho_k$ as
	\begin{equation}\label{Eq:ChainRule}
		\pd{f}{\rho_k} = \sum_{j= 1}^{Ne}\pd{f}{\bar{\rho_j}}\pd{\bar{\rho_j}}{\tilde{\rho_j}}\pd{\tilde{\rho_j}}{\rho_k},
	\end{equation}
	where $\pd{f}{\bar{\rho_j}}$ is evaluated using the adjoint-variable method (see Sec.~\ref{Subsec:Sensitivityanalysis}). We use the modified SIMP (Simplified Isotropic Material with Penalization) method to interpolate the Young's modulus of each FE using its physical design variable $\bar{\rho_j}$ as
	\begin{equation}\label{Eq:SIMPModel}
		E_1(\bar{\rho_j}) = E_0 + (\bar{\rho_j})^\zeta(E_1 -E_0), \qquad \bar{\rho_j}\in [0,\,1]
	\end{equation}
	where $E_1$ and $E_0$ are Young's moduli of the solid and void phases of an FE, respectively. The material contrast, i.e., $\frac{E_0}{E_1} = 10^{-6}$ is set, and the penalty factor $\zeta=3$ is used in order to steer the topology optimization towards a \textquoteleft0-1' solution. 
	
	\subsection{Robust formulation}
	The robust formulation is employed wherein three physical density fields, i.e.,  dilated $\bm{\bar{\rho}}^d$, intermediate (blueprint) $\bm{\bar{\rho}}^i$ and eroded $\bm{\bar{\rho}}^e$, are considered for the design domain \citep{wang2011projection}. Erosion and dilation are  morphological image operators, which can be used in TO for e.g. robustness and feature size control \citep{sigmund2007morphology}. Assuming uniform manufacturing errors, maximum and minimum manufacturing limits are indicated by the dilated and eroded designs respectively, whereas the intermediate (blueprint) ones denote the desired manufacturing limit. Here, 0.5$+\Delta\eta$, 0.5 and 0.5$-\Delta\eta$  in Eq.~\ref{Eq:projectionfilter} are used in place of $\eta$ to evaluate $\bm{\bar{\rho}}^e$, $\bm{\bar{\rho}}^i$ and $\bm{\bar{\rho}}^d$, respectively. The deviation $\Delta\eta \in[0,\,0.5]$ is a user defined parameter,  which in combination with the filter radius $r_\text{fil}$ determines the minimum length scale on the solid and void phases \citep{trillet2021analytical}. 
	
	The optimization problem is formulated as a min-max problem \citep{wang2011projection}   
	
	\begin{equation}\label{Eq:actualoptimization}
		\small
		\begin{rcases}
			\underset{\bm{\rho}}{\text{min}}:\text{max}
			: \left(f_0(\bar{\bm{\rho}}^d(\bm{\rho})),\,f_0(\bar{\bm{\rho}}^i(\bm{\rho})),\,f_0(\bar{\bm{\rho}}^e(\bm{\rho}))\right) \\
			\textit{s.t.} :\mathbf{A(\bm{\rho}}^l)\mathbf{p}(\bm{\rho}^l) = \mathbf{0},\,\qquad l = d,\,i,\,e \\
			\quad\mathbf{K(\bm{\rho}}^l)\mathbf{u}(\bm{\rho}^l) = \mathbf{F} = -\mathbf{D} \mathbf{p}(\bm{\rho}^l)\\
			\quad\mathbf{K(\bm{\rho}}^l)\mathbf{v}(\bm{\rho}^l) = \mathbf{F}_\mathrm{d}\\
			\quad V(\bar{\bm{\rho}}^d(\bm{\rho}))-V_d^*\le 0 \\
			\quad\mvect{0}\le\bm{\rho}\le\mvect{1}
		\end{rcases},
	\end{equation}
	where $f_0$ is a multi-criteria objective aimed at obtaining effective compliant mechanisms \citep{saxena2000optimal} defined by  $-\mu\frac{MSE}{SE}$, with $MSE = \trr{\mathbf{v}}\mvect{Ku}$ and $SE = \frac{1}{2}\trr{\mvect{u}}\mvect{Ku}$.  $MSE$ and $SE$ represent the mutual strain energy and strain energy of the mechanism, respectively. $\mu$ is the scaling factor used to scale the objective for optimizer compatibility.  Note that the multi-criteria objective proposed by \citet{frecker1997topological} finds an optimum trade-off between the flexibility and stiffness of the mechanisms. $V(\bar{\bm{\rho}}^d(\bm{\rho})) =\sum_{m=1}^{n_e} V_m\bar{{\rho}}^d_m$, where $V_m$ is the volume of $m^\text{th}$ element whose dilated density  is $\bar{{\rho}}^d_m$.  The volume constraint is imposed using the dilated design wherein the actual volume of the dilated design is updated after a specific number of optimization iterations such that the volume of the intermediate design becomes equal to the permitted one at the end of the optimization when the volume constraint becomes active~\citep{wang2011projection}. Further, $V_d^* = \frac{V_i^*}{{V(\bar{\bm{\rho}}^i(\bm{\rho}))}}V(\bar{\bm{\rho}}^d(\bm{\rho}))$, where $V_d^*$ denotes upper limit of the volume fraction of the dilated design, $V_i^*$ and $V(\bar{\bm{\rho}}^i(\bm{\rho}))=\sum_{m=1}^{n_e} V_m\bar{{\rho}}^i_m(\bm{\rho})$ are the prescribed and actual volumes of the intermediate design, respectively.

	The robust formulation (Eq.~\ref{Eq:actualoptimization}) requires solutions to three state equations pertaining to $\mathbf{u,\,p,\,v}$ fields and also furnishes three optimized designs with only one design vector $\bm{\rho}$.  Readers may refer to the paper by \citet{trillet2021analytical} for a complete discussion on the minimum feature size with the three-field design representation technique. The discreteness of the optimized solutions is measured using a gray scale indicator $M_\text{nd}$ defined as \citep{sigmund2007morphology}
	\begin{equation}
		M_\text{nd} = \frac{\displaystyle\sum_{e=1}^{n_{e}}4(\bar{\rho_e})(1-\bar{\rho_e})}{n_{e}},
	\end{equation}
	where $n_{e}$ is the total number of elements employed to discretize the design domain.
	
	\begin{figure*}
		\centering
		\begin{subfigure}[t]{0.30\textwidth}
			\includegraphics[scale=0.65]{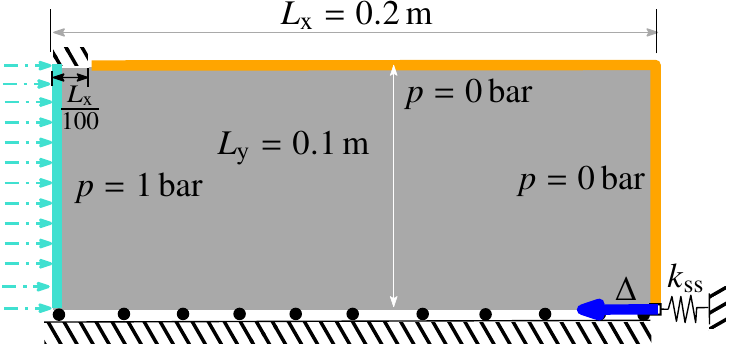}
			\caption{}
			\label{fig:inverter}
		\end{subfigure}
		\begin{subfigure}[t]{0.30\textwidth}
			\includegraphics[scale=0.65]{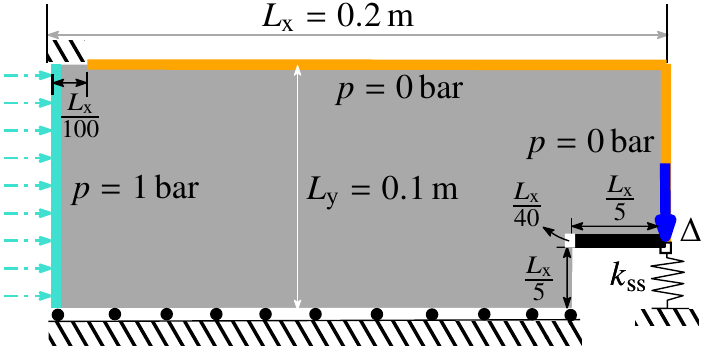}
			\caption{}
			\label{fig:gripper}
		\end{subfigure}
		\begin{subfigure}[t]{0.30\textwidth}
			\includegraphics[scale=0.65]{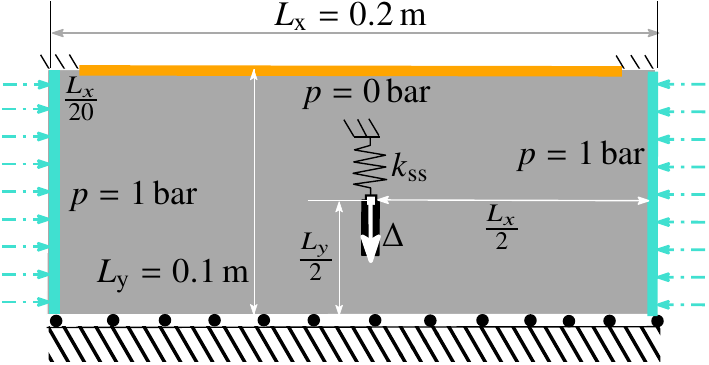}
			\caption{}
			\label{fig:contractor}
		\end{subfigure}
		\caption{Symmetric half design domains of the mechanisms. (\subref{fig:inverter}) Inverter design domain, (\subref{fig:gripper}) Gripper design domain and (\subref{fig:contractor}) Contractor design domain. The dimensions of the design domains are $L_x =\SI{0.2}{\meter}$ and  $L_y = \SI{0.1}{\meter}$, where $L_x$ and $L_y$ indicate dimensions in $x-$ and $y-$directions, respectively. For the inverter and gripper mechanisms, the input pressure load is applied on the left edge, and zero pressure is applied at the other edges except the symmetric boundary. The contractor mechanism is actuated from both the left and right edges. Symmetric boundaries and fixed boundaries are also depicted. Workpiece stiffnesses are represented via the output springs with stiffnesses $k_\text{ss}$. In gripper mechanism domain, a non-design void region having area $\frac{L_x}{5}\times\frac{L_x}{5}$ and a solid region having area $\frac{L_x}{5}\times\frac{L_x}{40}$ are used at the right lower part. A solid non-design domain of size $\frac{L_x}{40}\times \frac{L_y}{4}$ is present in the middle of the symmetric half domain of the contractor mechanism.} \label{fig:DesignDomain}
	\end{figure*}
	
	\begin{table*}
		\centering
		\caption{Various parameters used in this paper.}\label{Table:T1}
		\begin{threeparttable}
			\begin{tabular}{ l c c }
				\hline
				\hline
				\textbf{Parameter} & \textbf{Notation} & \textbf{Value}  \\ \hline
				Young's modulus of actual material & $E_1$ & $\SI{3e9}{\newton\per\square\meter}$ \rule{0pt}{3ex}\\ 
				Poisson's ratio & $\nu$ & $0.40$\\ 
				Out-of-plane thickness & $t$ & $\SI{0.001}{\meter}$\\ 
				Penalization  &$\zeta$ &$3$ \rule{0pt}{3ex}   \\
				Young's modulus of a void FE ($\rho=0$) & $E_0$ &$E_1 \times 10^{-6} \si{\newton\per\square\meter}$ \\
				External move limit\tnote{2} & $\Delta \bm{\rho}$ & 0.1 per iteration\\
				Input pressure load &$p_\mathrm{in}$ & $\SI{1e5}{\newton\per\square\meter}$ \rule{0pt}{3ex}\\
				$K(\bm{\rho})$ step location	& 	$\eta_k$ 	& 0.3 \rule{0pt}{3ex}\\
				$K(\bm{\rho})$ slope	at step	& 	$\beta_k$	& 10\\
				$H(\bm{\rho})$ step location	& 	$\eta_h$ 	& 0.2\\
				$H(\bm{\rho})$ slope	at step	& 	$\beta_h$	& 10\\ 
				Flow coefficient of a void FE	&   $k_\mathrm{v}$  & $\SI{1}{\meter\tothe{4}\per\newton\per\second}$\\
				Flow coefficient of a solid FE &   $k_\mathrm{s} $  & $k_\mathrm{v}\times\SI{e-7}{\meter\tothe{4}\per\newton\per\second}$\\
				Drainage from solid	&   $h_\mathrm{s} $  & $\left(\frac{\ln{r}}{\Delta s}\right)^2 k_\mathrm{s}$ \\
				Remainder of input pressure at $\Delta s$ &r& 0.1\\\hline\hline
			\end{tabular}
			\begin{tablenotes}
				\item[2] The external move limit is used to update \texttt{xminvec} and \texttt{xmaxvec} of the MMA outside of the \texttt{mmasub} function call.
			\end{tablenotes}
		\end{threeparttable}
	\end{table*}
	
	\begin{figure*}
		\begin{subfigure}[t]{0.30\textwidth}
			\centering
			\includegraphics[scale=.40]{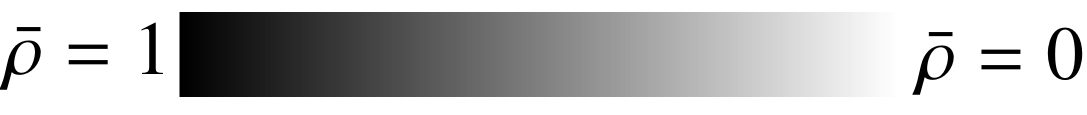}
			\caption{}
			\label{fig:colorMaterial}
		\end{subfigure}
		\begin{subfigure}[t]{0.30\textwidth}
			\centering
			\includegraphics[scale=.40]{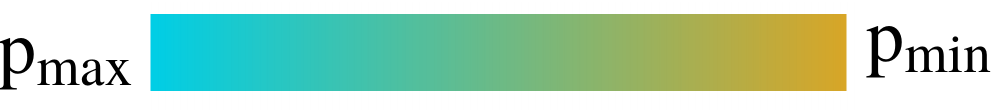}
			\caption{}
			\label{fig:colorPressure}
		\end{subfigure}
		\begin{subfigure}[t]{0.30\textwidth}
			\centering
			\includegraphics[scale=.40]{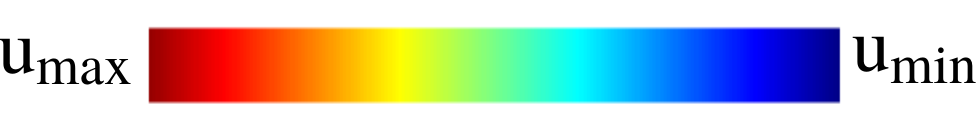}
			\caption{}
			\label{fig:colorDisplacement}
		\end{subfigure}
		\caption{Color schemes employed in this paper to plot matrial, pressure and displacement fields are shown in (\subref{fig:colorMaterial}), (\subref{fig:colorPressure}) and (\subref{fig:colorDisplacement}) respectively. p$_\text{max}=\SI{1}{\bar}$ and p$_\text{min}=\SI{0}{\bar}$ represent the maximum and minimum values of the pressue load. Maximum and minimum values of the magnitude of displacement field are indicated via u$_\text{min} =0$ and u$_\text{max}$ respectively.} \label{fig:colorschem}
	\end{figure*}
	
	\subsection{Sensitivity analysis}\label{Subsec:Sensitivityanalysis}
	We use the Method of Moving Asymptotes (MMA)~\citep{svanberg1987method}, a gradient-based optimizer, for solving the optimization problem~(Eq.~\ref{Eq:actualoptimization}). A standard setting available in the MMA optimizer is used to solve the min-max optimization problem. The Lagrangian $\mathcal{L}$ using the objective function and constraints can be written as
	\begin{equation}\label{Eq:augmentedperformance}
		\mathcal{L} = f_0(\bar{\bm{\rho}}) + \trr{\bm{\lambda}}_1 \left(\mathbf{Ku +{H p}}\right) + \trr{\bm{\lambda}}_2 ( \mathbf{Ap}) + \trr{\bm{\lambda}}_3 (\mathbf{Kv-F_\mathrm{d}}) + \Lambda \left(V-V_d^*\right),
	\end{equation}
	where $\bm{\lambda}_i|_{i= 1,\,2,\,3}$ and $\Lambda$ are the Lagrange multipliers. Using the adjoint equations corresponding to Eq.~\ref{Eq:augmentedperformance}, i.e., $\frac{\partial\mathcal{L}}{\partial \mathbf{u}} = 0,\,\frac{\partial\mathcal{L}}{\partial \mathbf{p}} = 0,\,\text{and}\,\frac{\partial\mathcal{L}}{\partial \mathbf{v}} = 0$, one finds the Lagrange multipliers $\bm{\lambda}_1,\,\bm{\lambda}_2$ and $\bm{\lambda_3}$ as \citep{kumar2020topology} 
	
	\begin{equation}\label{Eq:lagrangemultipliers}
		\begin{rcases}
			\trr{\bm{\lambda}}_1  &= -\pd{f_0(\mathbf{u},\, \mathbf{v},\,\bm{\rho})}{\mathbf{u}} \inv{\mathbf{K}}\\
			\trr{\bm{\lambda}}_2  & = -\trr{\bm{\lambda}}_1 \mathbf{H}\inv{\mathbf{A}}\\
			\trr{\bm{\lambda}}_3  &= -\pd{f_0(\mathbf{u},\, \mathbf{v},\,\bm{\rho})}{\mathbf{v}} \inv{\mathbf{K}}
		\end{rcases}.
	\end{equation} 	
	
	\begin{figure*}
		\centering
		\includegraphics[scale=2]{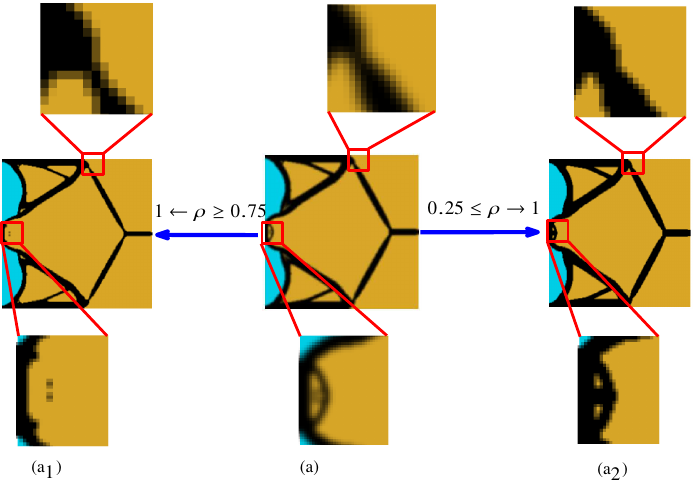}
		\caption{An optimized pressure-actuated inverter mechanism obtained using the method presented in \cite{kumar2020topology} is displayed in (a). The optimized design contains thin flexure regions surrounded by gray elements that are depicted in insets. Using the two different thresholds, the approximated designs are displayed in (a$_1$) and (a$_2$).} \label{fig:Inverterwithoutrobust}
	\end{figure*}

	\begin{figure*}
		\begin{subfigure}[t]{0.30\textwidth}
			\centering
			\includegraphics[scale=.4]{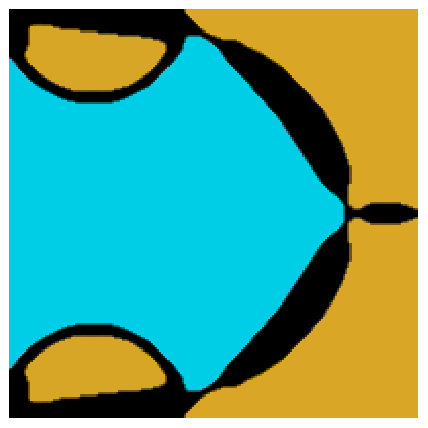}
			\caption{Eroded design, $\Delta = \SI{-0.33}{\milli \meter}$}
				\footnotesize$\Delta\eta=0.15$:		$M_\text{nd}=0.16\%$,\,$V_f=0.178$
			\label{fig:ricase1e}
		\end{subfigure}
		\begin{subfigure}[t]{0.30\textwidth}
			\centering
			\includegraphics[scale=.4]{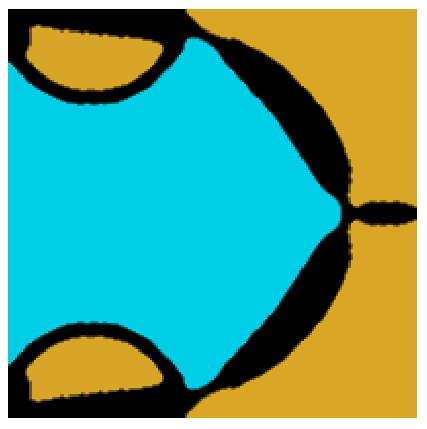}
			\caption{Intermediate design, $\Delta = \SI{-0.29}{\milli \meter}$}
			$M_\text{nd}=0.47\%$,\,$V_f=0.20$
			\label{fig:ricase1i}
		\end{subfigure}
		\begin{subfigure}[t]{0.30\textwidth}
			\centering
			\includegraphics[scale=.4]{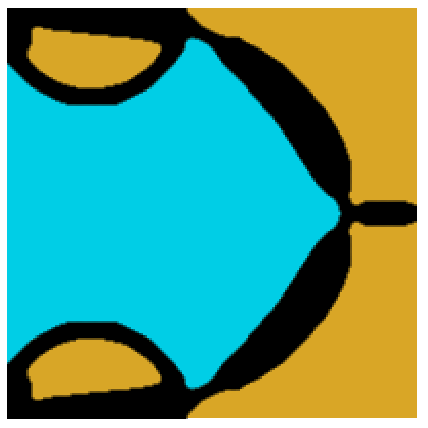}
			\caption{Dilated design, $\Delta = \SI{-0.24}{\milli \meter}$}
			$M_\text{nd}=0.22\%$,\,$V_f=0.22$
			\label{fig:ricase1d}
		\end{subfigure}
		\begin{subfigure}[t]{0.30\textwidth}
			\centering
			\includegraphics[scale=.4]{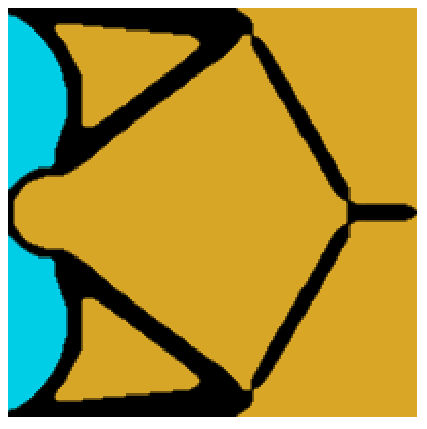}
			\caption{Eroded design, $\Delta = \SI{-0.25}{\milli \meter}$}
			$\Delta\eta=0.05$:	$M_\text{nd}=0.14\%$,\,$V_f=0.19$
			\label{fig:ricase2e}
		\end{subfigure}
		\begin{subfigure}[t]{0.30\textwidth}
			\centering
			\includegraphics[scale=.4]{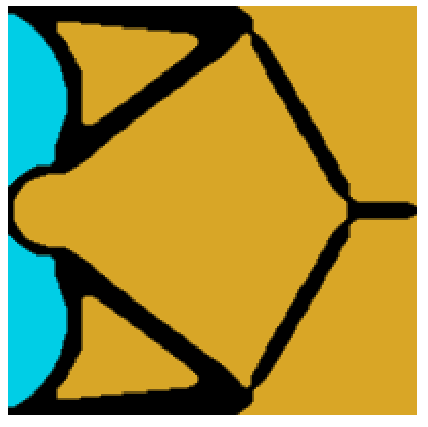}
			\caption{Intermediate design, $\Delta = \SI{-0.240}{\milli \meter}$}
			$M_\text{nd}=0.29\%$,\,$V_f=0.20$
			\label{fig:ricase2i}
		\end{subfigure}
		\begin{subfigure}[t]{0.3\textwidth}
			\centering
			\includegraphics[scale=.4]{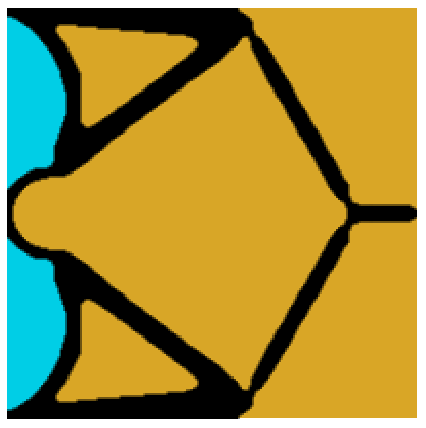}
			\caption{Dilated design, $\Delta = \SI{-0.23}{\milli \meter}$}
			$M_\text{nd}=0.27\%$,\,$V_f=0.21$
			\label{fig:ricase2d}
		\end{subfigure}
		\begin{subfigure}[t]{0.3\textwidth}
			\centering
			\includegraphics[scale=0.4]{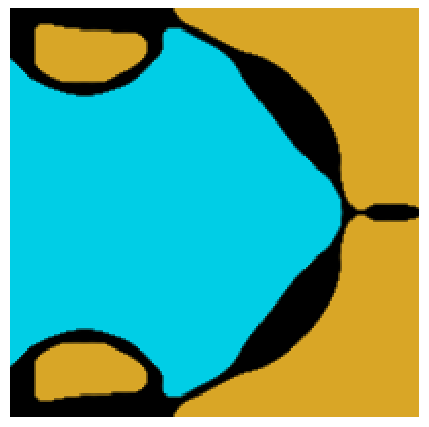}
			\caption{Eroded design, $\Delta = \SI{-0.36}{\milli \meter}$}
			$\Delta\eta=0.15$:  $M_\text{nd}=0.37\%$,\,$V_f=0.16$
			\label{fig:ricase3e}
		\end{subfigure}
		\begin{subfigure}[t]{0.3\textwidth}
			\centering
			\includegraphics[scale=.4]{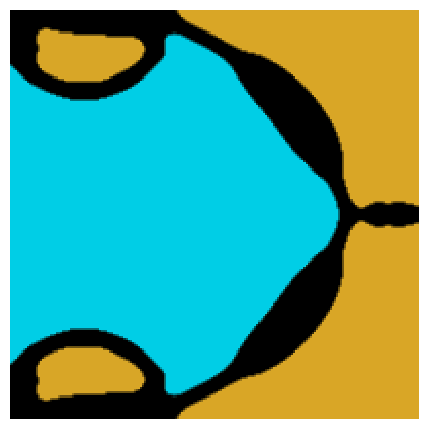}
			\caption{Intermediate design, $\Delta = \SI{-0.29}{\milli \meter}$}
			$M_\text{nd}=0.28\%$,\,$V_f=0.020$
			\label{fig:ricase3i}
		\end{subfigure}
		\begin{subfigure}[t]{0.3\textwidth}
			\centering
			\includegraphics[scale=.4]{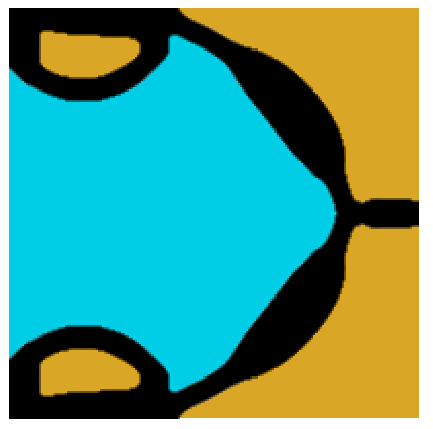}
			\caption{Dilated design, $\Delta = \SI{-0.21}{\milli \meter}$}
			$M_\text{nd}=0.3\%$,\,$V_f=0.24$
			\label{fig:ricase3d}
		\end{subfigure}
		\begin{subfigure}[t]{0.3\textwidth}
			\centering
			\includegraphics[scale=0.4]{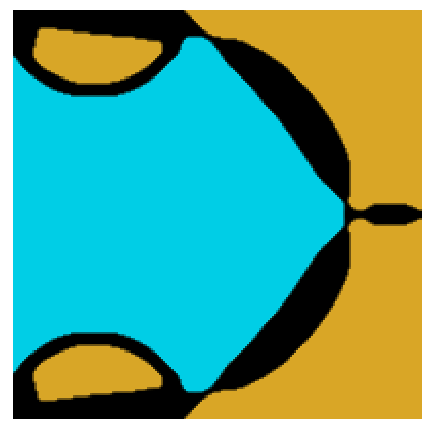}
			\caption{Eroded design, $\Delta = \SI{-0.32}{\milli \meter}$}
			$\Delta\eta=0.05$: $M_\text{nd}=0.28\%$,\,$V_f=0.19$
			\label{fig:ricase4e}
		\end{subfigure}
		\begin{subfigure}[t]{0.3\textwidth}
			\centering
			\includegraphics[scale=.4]{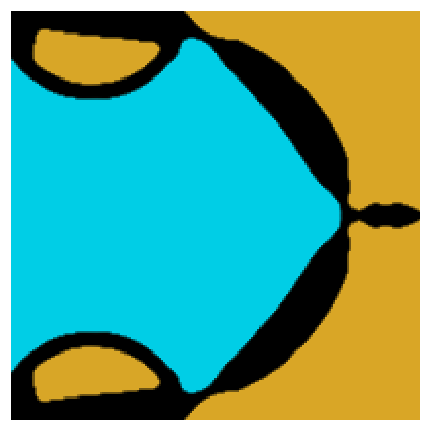}
			\caption{Intermediate design, $\Delta = \SI{-0.29}{\milli \meter}$}
			$M_\text{nd}=0.62\%$,\,$V_f=0.20$
			\label{fig:ricase4i}
		\end{subfigure}
		\begin{subfigure}[t]{0.3\textwidth}
			\centering
			\includegraphics[scale=.4]{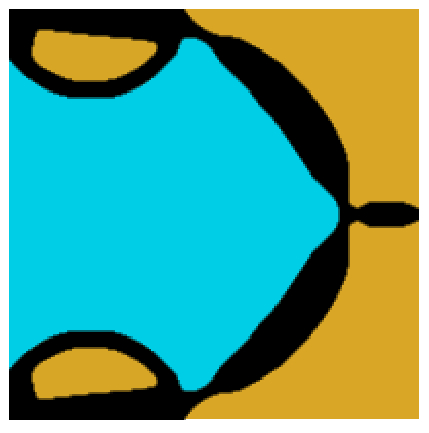}
			\caption{Dilated design, $\Delta = \SI{-0.25}{\milli \meter}$}
			$M_\text{nd}=0.23\%$,\,$V_f=0.21$
			\label{fig:ricase4d}
		\end{subfigure}
		\caption{The pressure-actuated inverter mechanisms. Filter radius $5.4h$ is used for the optimized results shown in rows 1 and 2, whereas for the results displayed in rows 3 and 4, it is set to $8.4h$.  Note $h =\min\left(\frac{L_x}{N_\text{ex}},\,\frac{L_y}{N_\text{ey}}\right)$. We refer the inverter mechanisms in the first, second, third and fourth rows as IV1 Pa-CMs, IV2 Pa-CMs, IV3 Pa-CMs and IV4 Pa-CMs respectively.} \label{fig:robustinverters}
	\end{figure*}
	
	\begin{figure*}
		\begin{subfigure}[t]{0.15\textwidth}
			\centering
			\includegraphics[scale = 0.20]{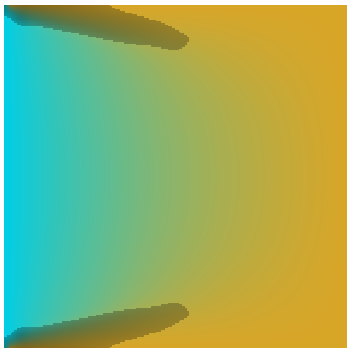}
			\caption{Iteration 5}
			\label{fig:IV1_5MMA}
		\end{subfigure}
		\begin{subfigure}[t]{0.15\textwidth}
			\centering
			\includegraphics[scale = 0.20]{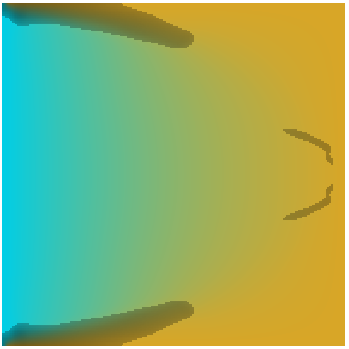}
			\caption{Iteration 6}
			\label{fig:IV1_6MMA}
		\end{subfigure}
		\begin{subfigure}[t]{0.15\textwidth}
			\centering
			\includegraphics[scale = 0.20]{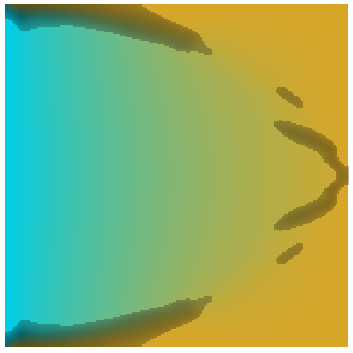}
			\caption{Iteration 7}
			\label{fig:IV1_7MMA}
		\end{subfigure}
		\begin{subfigure}[t]{0.15\textwidth}
			\centering
			\includegraphics[scale = 0.20]{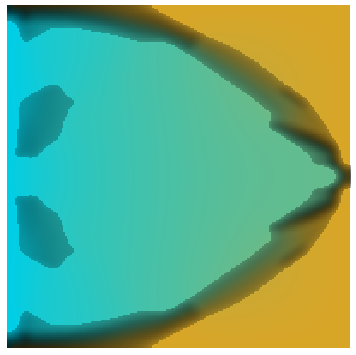}
			\caption{Iteration 10}
			\label{fig:IV1_10MMA}
		\end{subfigure}
		\begin{subfigure}[t]{0.15\textwidth}
			\centering
			\includegraphics[scale = 0.20]{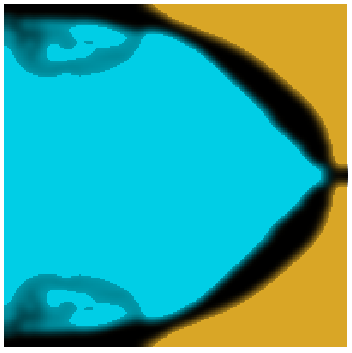}
			\caption{Iteration 20}
			\label{fig:IV1_20MMA}
		\end{subfigure}
		\begin{subfigure}[t]{0.15\textwidth}
			\centering
			\includegraphics[scale = 0.20]{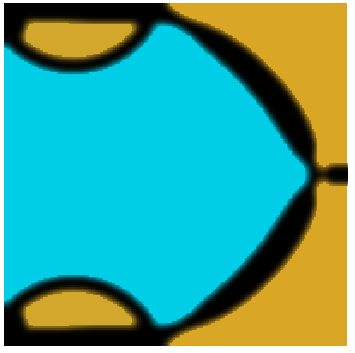}
			\caption{Iteration 50}
			\label{fig:IV1_50MMA}
		\end{subfigure}
		\caption{IV1 Pa-CM designs at different MMA iterations.}	\label{fig:IV1_MMA}
	\end{figure*}
	
	\begin{figure*}
		\begin{subfigure}[t]{0.15\textwidth}
			\centering
			\includegraphics[scale = 0.20]{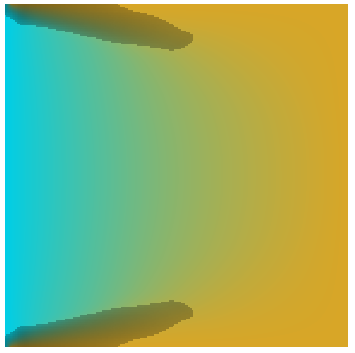}
			\caption{Iteration 5}
			\label{fig:IV2_5MMA}
		\end{subfigure}
		\begin{subfigure}[t]{0.15\textwidth}
			\centering
			\includegraphics[scale = 0.20]{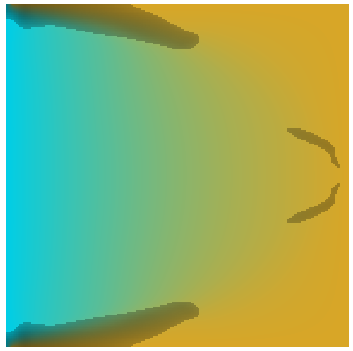}
			\caption{Iteration 6}
			\label{fig:IV2_6MMA}
		\end{subfigure}
		\begin{subfigure}[t]{0.15\textwidth}
			\centering
			\includegraphics[scale = 0.20]{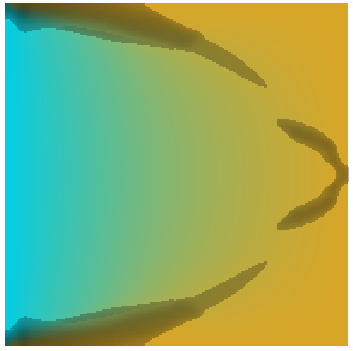}
			\caption{Iteration 7}
			\label{fig:IV2_7MMA}
		\end{subfigure}
		\begin{subfigure}[t]{0.15\textwidth}
			\centering
			\includegraphics[scale = 0.20]{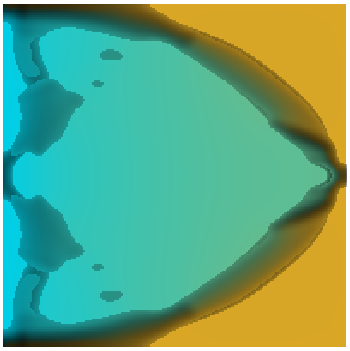}
			\caption{Iteration 10}
			\label{fig:IV2_10MMA}
		\end{subfigure}
		\begin{subfigure}[t]{0.15\textwidth}
			\centering
			\includegraphics[scale = 0.20]{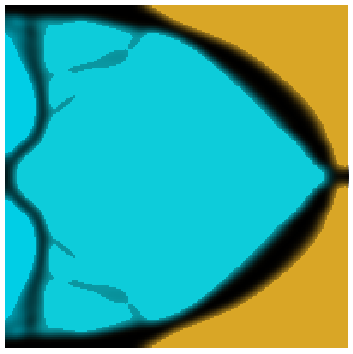}
			\caption{Iteration 20}
			\label{fig:IV2_20MMA}
		\end{subfigure}
		\begin{subfigure}[t]{0.15\textwidth}
			\centering
			\includegraphics[scale = 0.20]{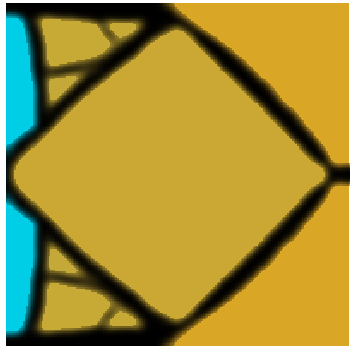}
			\caption{Iteration 50}
			\label{fig:IV2_50MMA}
		\end{subfigure}
		\caption{IV2 Pa-CM designs at different MMA iterations.}	\label{fig:IV2_MMA}
	\end{figure*}
	
	
	\begin{figure*}
		\begin{subfigure}[t]{0.30\textwidth}
			\centering
			\includegraphics[scale=.4]{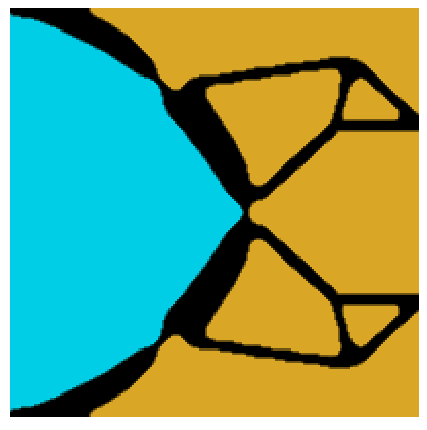}
			\caption{Eroded design, $\Delta = \SI{-0.22}{\milli \meter}$}
			$\Delta\eta=0.15$:		$M_\text{nd}=0.21\%$,\,$V_f=0.17$
			\label{fig:rgcase1e}
		\end{subfigure}
		\begin{subfigure}[t]{0.30\textwidth}
			\centering
			\includegraphics[scale=.4]{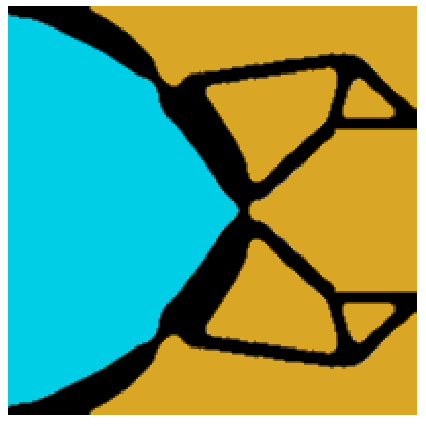}
			\caption{Intermediate design, $\Delta = \SI{-0.19}{\milli \meter}$}
			$M_\text{nd}=0.38\%$,\,$V_f=0.20$
			\label{fig:rgcase1i}
		\end{subfigure}
		\begin{subfigure}[t]{0.30\textwidth}
			\centering
			\includegraphics[scale=.4]{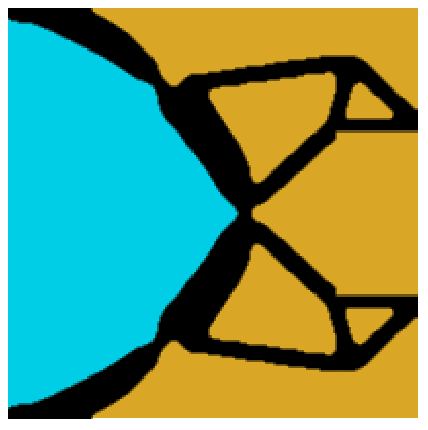}
			\caption{Dilated design, $\Delta = \SI{-0.16}{\milli \meter}$}
			$M_\text{nd}=0.12\%$,\,$V_f=0.23$
			\label{fig:rgcase1d}
		\end{subfigure}
		\begin{subfigure}[t]{0.30\textwidth}
			\centering
			\includegraphics[scale=.4]{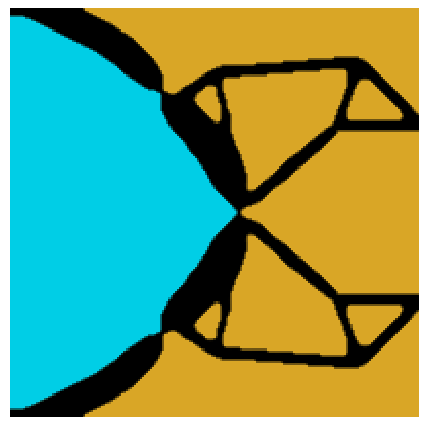}
			\caption{Eroded design, $\Delta = \SI{-0.21}{\milli \meter}$}
			$\Delta\eta=0.05$:	$M_\text{nd}=0.26\%$,\,$V_f=0.19$
			\label{fig:rgcase2e}
		\end{subfigure}
		\begin{subfigure}[t]{0.30\textwidth}
			\centering
			\includegraphics[scale=.4]{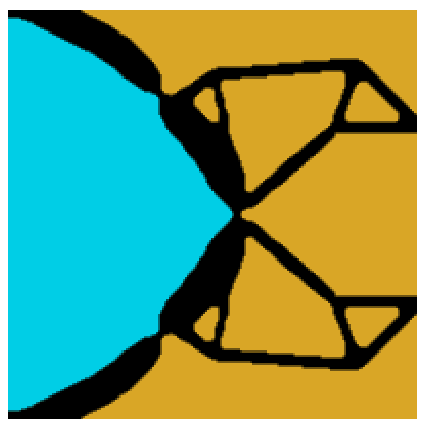}
			\caption{Intermediate design, $\Delta = \SI{-0.19}{\milli \meter}$}
			$M_\text{nd}=0.26\%$,\,$V_f=0.20$
			\label{fig:rgcase2i}
		\end{subfigure}
		\begin{subfigure}[t]{0.3\textwidth}
			\centering
			\includegraphics[scale=.4]{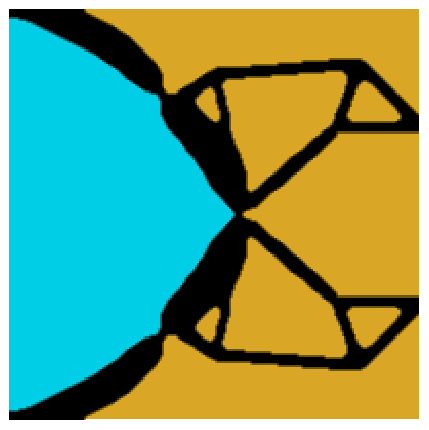}
			\caption{Dilated design, $\Delta = \SI{-0.18}{\milli \meter}$}
			$M_\text{nd}=0.34\%$,\,$V_f=0.21$
			\label{fig:rgcase2d}
		\end{subfigure}
		\begin{subfigure}[t]{0.3\textwidth}
			\centering
			\includegraphics[scale=0.4]{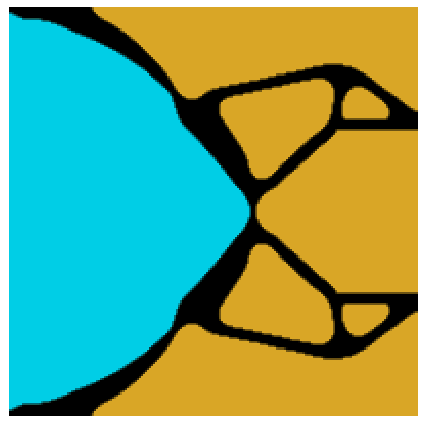}
			\caption{Eroded design, $\Delta = \SI{-0.25}{\milli \meter}$}
			$\Delta\eta=0.15$:  $M_\text{nd}=0.54\%$,\,$V_f=0.16$
			\label{fig:rgcase3e}
		\end{subfigure}
		\begin{subfigure}[t]{0.3\textwidth}
			\centering
			\includegraphics[scale=.4]{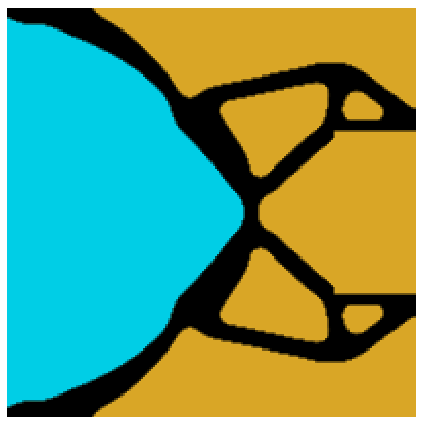}
			\caption{Intermediate design, $\Delta = \SI{-0.1896}{\milli \meter}$}
			$M_\text{nd}=0.51\%$,\,$V_f=0.020$
			\label{fig:rgcase3i}
		\end{subfigure}
		\begin{subfigure}[t]{0.3\textwidth}
			\centering
			\includegraphics[scale=.4]{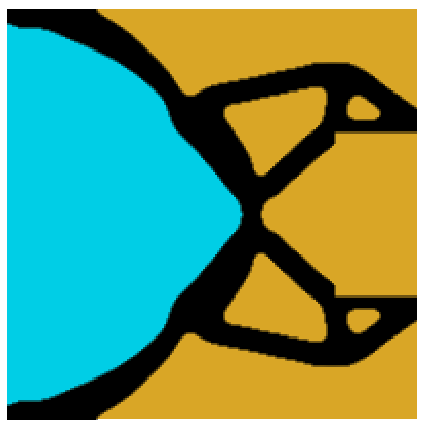}
			\caption{Dilated design, $\Delta = \SI{-0.1435}{\milli \meter}$}
			$M_\text{nd}=0.53\%$,\,$V_f=0.24$
			\label{fig:rgcase3d}
		\end{subfigure}
		\begin{subfigure}[t]{0.3\textwidth}
			\centering
			\includegraphics[scale=0.4]{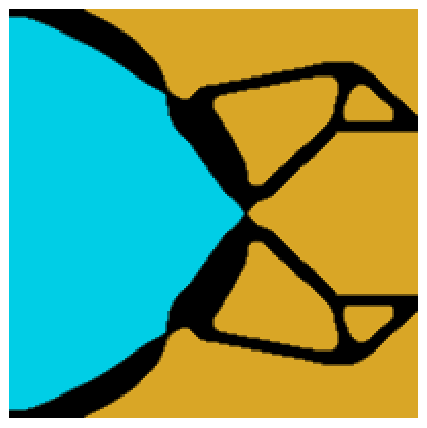}
			\caption{Eroded design, $\Delta = \SI{-0.23}{\milli \meter}$}
			$\Delta\eta=0.05$: $M_\text{nd}=0.37\%$,\,$V_f=0.18$
			\label{fig:rgcase4e}
		\end{subfigure}
		\begin{subfigure}[t]{0.3\textwidth}
			\centering
			\includegraphics[scale=.4]{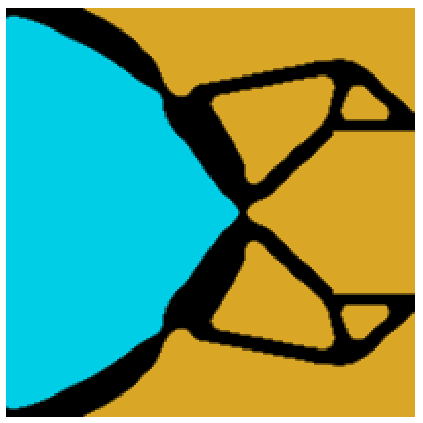}
			\caption{Intermediate design, $\Delta = \SI{-0.20}{\milli \meter}$}
			$M_\text{nd}=0.77\%$,\,$V_f=0.20$
			\label{fig:rgcase4i}
		\end{subfigure}
		\begin{subfigure}[t]{0.3\textwidth}
			\centering
			\includegraphics[scale=.4]{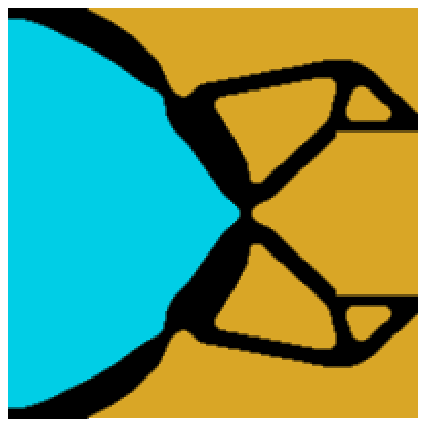}
			\caption{Dilated design, $\Delta = \SI{-0.18}{\milli \meter}$}
			$M_\text{nd}=0.32\%$,\,$V_f=0.22$
			\label{fig:rgcase4d}
		\end{subfigure}
		\caption{The robust pressure-actuated gripper mechanisms. Filter radius $5.4h$ is used for the optimized results displayed in rows 1 and 2, whereas for the results shown in rows 3 and 4 are obtained with filter radius $8.4h$.  Note $h =\min\left(\frac{L_x}{N_\text{ex}},\,\frac{L_y}{N_\text{ey}}\right)$.  We refer the gripper mechanisms in the first, second, third and fourth rows as GP1 Pa-CMs, GP2 Pa-CMs, GP3 Pa-CMs and GP4 Pa-CMs respectively.} \label{fig:robustgripper}
	\end{figure*}
	
	
	\begin{figure*}
		\begin{subfigure}[t]{0.30\textwidth}
			\centering
			\includegraphics[scale=.4]{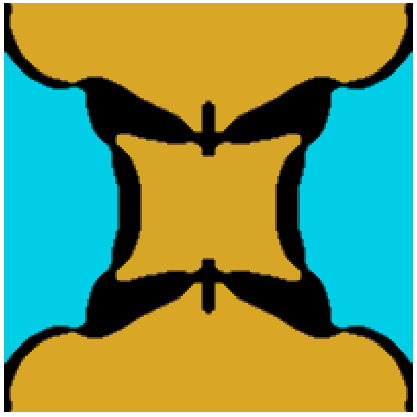}
			\caption{Eroded design, $\Delta = \SI{-0.13}{\milli \meter}$}
			$\Delta\eta=0.15$:		$M_\text{nd}=0.26\%$,\,$V_f=0.173$
			\label{fig:rccase1e}
		\end{subfigure}
		\begin{subfigure}[t]{0.30\textwidth}
			\centering
			\includegraphics[scale=.4]{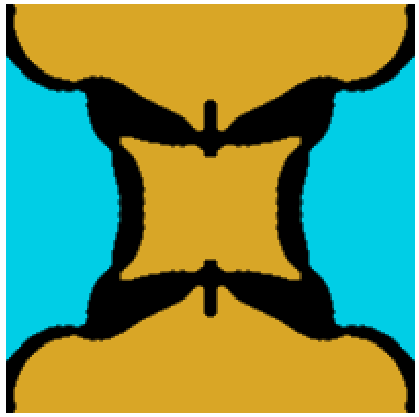}
			\caption{Intermediate design, $\Delta = \SI{-0.09}{\milli \meter}$}
			$M_\text{nd}=0.65\%$,\,$V_f=0.20$
			\label{fig:rccase1i}
		\end{subfigure}
		\begin{subfigure}[t]{0.30\textwidth}
			\centering
			\includegraphics[scale=.4]{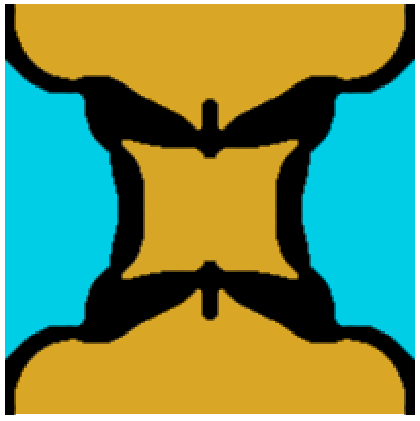}
			\caption{Dilated design, $\Delta = \SI{-0.07}{\milli \meter}$}
			$M_\text{nd}=0.27\%$,\,$V_f=0.23$
			\label{fig:rccase1d}
		\end{subfigure}
		\begin{subfigure}[t]{0.30\textwidth}
			\centering
			\includegraphics[scale=.4]{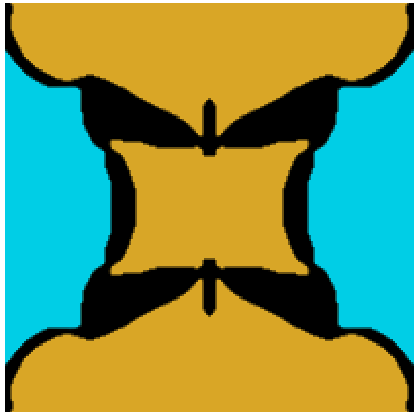}
			\caption{Eroded design, $\Delta = \SI{-0.076}{\milli \meter}$}
			$\Delta\eta=0.05$:	$M_\text{nd}=0.21\%$,\,$V_f=0.19$
			\label{fig:rccase2e}
		\end{subfigure}
		\begin{subfigure}[t]{0.30\textwidth}
			\centering
			\includegraphics[scale=.4]{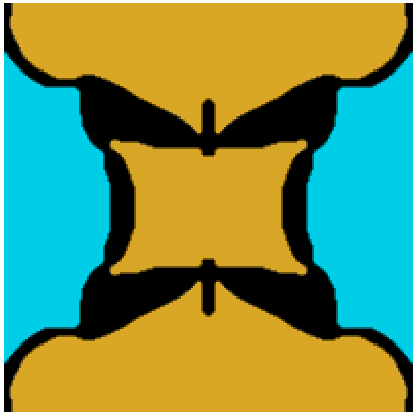}
			\caption{Intermediate design, $\Delta = \SI{-0.065}{\milli \meter}$}
			$M_\text{nd}=0.30\%$,\,$V_f=0.20$
			\label{fig:rccase2i}
		\end{subfigure}
		\begin{subfigure}[t]{0.3\textwidth}
			\centering
			\includegraphics[scale=.4]{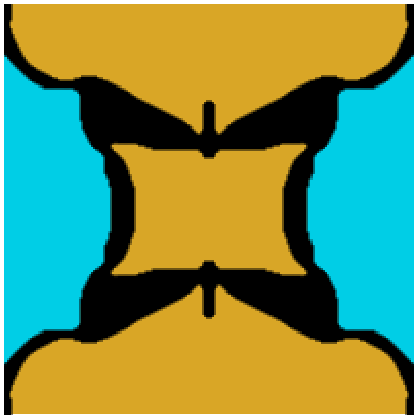}
			\caption{Dilated design, $\Delta = \SI{-0.060}{\milli \meter}$}
			$M_\text{nd}=0.22\%$,\,$V_f=0.21$
			\label{fig:rccase2d}
		\end{subfigure}
		\begin{subfigure}[t]{0.3\textwidth}
			\centering
			\includegraphics[scale=0.4]{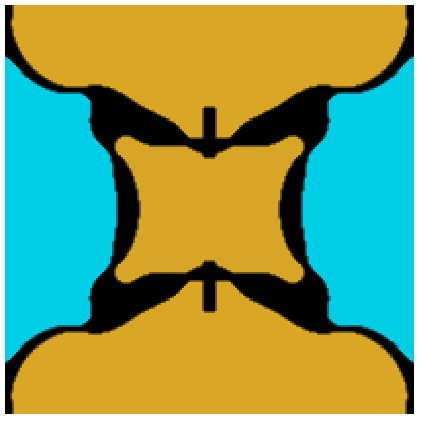}
			\caption{Eroded design, $\Delta = \SI{-0.14}{\milli \meter}$}
			$\Delta\eta=0.15$:  $M_\text{nd}=0.46\%$,\,$V_f=0.15$
			\label{fig:rccase3e}
		\end{subfigure}
		\begin{subfigure}[t]{0.3\textwidth}
			\centering
			\includegraphics[scale=.4]{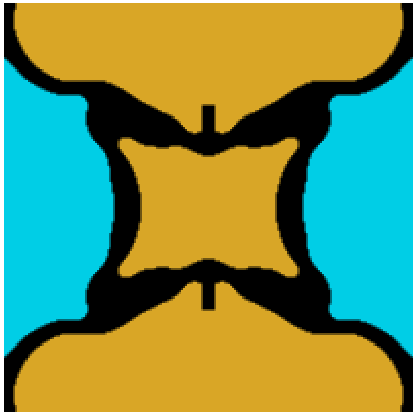}
			\caption{Intermediate design, $\Delta = \SI{-0.078}{\milli \meter}$}
			$M_\text{nd}=0.30\%$,\,$V_f=0.020$
			\label{fig:rccase3i}
		\end{subfigure}
		\begin{subfigure}[t]{0.3\textwidth}
			\centering
			\includegraphics[scale=.4]{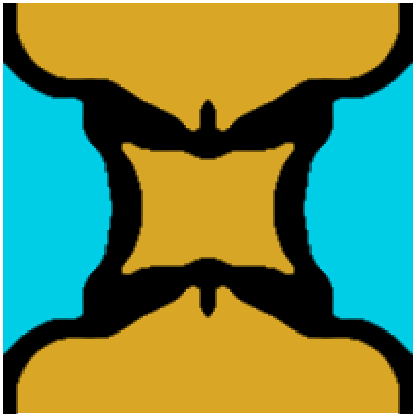}
			\caption{Dilated design, $\Delta = \SI{-0.058}{\milli \meter}$}
			$M_\text{nd}=0.41\%$,\,$V_f=0.25$
			\label{fig:rccase3d}
		\end{subfigure}
		\begin{subfigure}[t]{0.3\textwidth}
			\centering
			\includegraphics[scale=0.4]{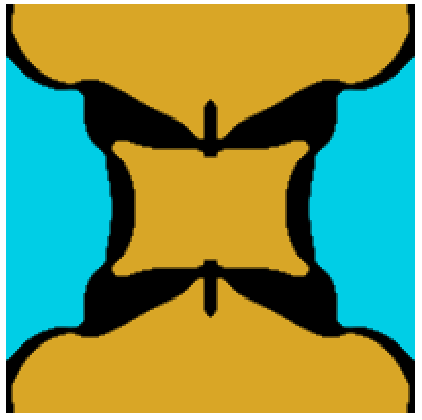}
			\caption{Eroded design, $\Delta = \SI{-0.083}{\milli \meter}$}
			$\Delta\eta=0.05$: $M_\text{nd}=0.33\%$,\,$V_f=0.18$
			\label{fig:rccase4e}
		\end{subfigure}
		\begin{subfigure}[t]{0.3\textwidth}
			\centering
			\includegraphics[scale=.4]{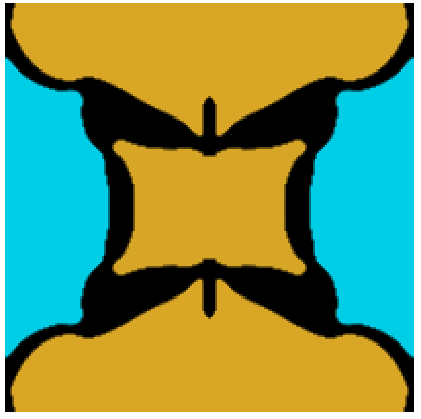}
			\caption{Intermediate design, $\Delta = \SI{-0.063}{\milli \meter}$}
			$M_\text{nd}=0.68\%$,\,$V_f=0.20$
			\label{fig:rccase4i}
		\end{subfigure}
		\begin{subfigure}[t]{0.3\textwidth}
			\centering
			\includegraphics[scale=.4]{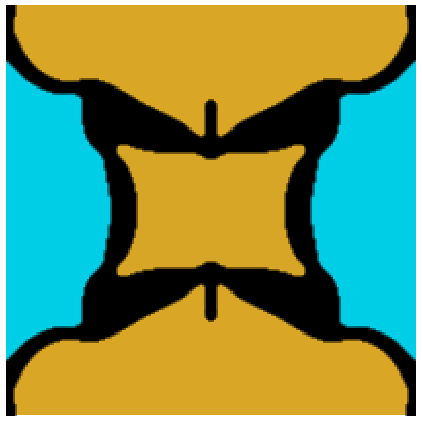}
			\caption{Dilated design, $\Delta = \SI{-0.052}{\milli \meter}$}
			$M_\text{nd}=0.4\%$,\,$V_f=0.22$
			\label{fig:rccase4d}
		\end{subfigure}
		\caption{The pressure-actuated contractor mechanisms. Filter radius $5.4h$ is used for the optimized results shown in rows 1 and 2, whereas for the results displayed in rows 3 and 4, it is set to $8.4h$. Note $h =\min\left(\frac{L_x}{N_\text{ex}},\,\frac{L_y}{N_\text{ey}}\right)$. We refer the inverter mechanisms in the first, second, third and fourth rows as CT1 Pa-CMs, CT2 Pa-CMs, CT3 Pa-CMs and CT4 Pa-CMs respectively.} \label{fig:robustcontractor}
	\end{figure*}
	
	The design equation corresponding to Eq.~\ref{Eq:augmentedperformance} is
	\begin{equation}\label{Eq:sensitivityofobjective1}
		\begin{aligned}
			\frac{\partial \mathcal{L}}{\partial \bar{\bm{\rho}}} = &\pd{f_0}{\bar{\bm{\rho}}} + \trr{\bm{\lambda}}_1\pd{\mathbf{K}}{\bar{\bm{\rho}}}\mathbf{u} + \trr{\bm{\lambda}}_2\pd{\mathbf{A}}{\bar{\bm{\rho}}}\mathbf{p} +  \trr{\bm{\lambda}}_3\pd{\mathbf{K}}{\bar{\bm{\rho}}}\mathbf{v} + \Lambda \frac{\partial \left(V(\bar{\bm{\rho}}^d(\bm{\rho}))-V_d^*\right)}{\partial \bar{\bm{\rho}}}\\
			=&\underbrace{\pd{f_0}{\bar{\bm{\rho}}} + \trr{\bm{\lambda}}_1\pd{\mathbf{K}}{\bar{\bm{\rho}}}\mathbf{u} + \trr{\bm{\lambda}}_2\pd{\mathbf{A}}{\bar{\bm{\rho}}}\mathbf{p} +  \trr{\bm{\lambda}}_3\pd{\mathbf{K}}{\bar{\bm{\rho}}}\mathbf{v}}_{\Theta} + \Lambda \frac{\partial \left(V(\bar{\bm{\rho}}^d(\bm{\rho}))\right)}{\partial \bar{\bm{\rho}}},
		\end{aligned}
	\end{equation} 
	with complementarity condition $\Lambda \left(V(\bar{\bm{\rho}}^d(\bm{\rho}))-V_d^*\right) = 0,\,\Lambda\ge0$.
	Using $f_0 = -\mu\frac{MSE}{SE}$ and in view of Eq.~\ref{Eq:lagrangemultipliers}, $\Theta = \Theta_1 + \Theta_2$ transpires as 
	\begin{equation}\label{Eq:sensitivityofobjective2}
		\begin{aligned}
			\Theta_1  =  \mu\left[\trr{\mathbf{u}}\pd{\mathbf{K}}{\bar{\bm{\rho}}}\left(-\mathbf{u}\frac{MSE}{2(SE)^2} + \frac{\mathbf{v}}{SE}\right)\right],
		\end{aligned}
	\end{equation}
	\begin{equation}
		\begin{aligned}
			\Theta_2 =  \mu\left[\left(\frac{MSE}{(SE)^2}\trr{\mathbf{u}}  + \frac{-\trr{\mathbf{v}}}{SE} \right)\mathbf{H}\inv{\mathbf{A}}\pd{\mathbf{A}}{\bar{\bm{\rho}}}\mathbf{p}\right].
		\end{aligned}
	\end{equation}
	where $\Theta_1$ and $\Theta_2$ represent objective and load sensitivities respectively.
	Now, using Eq.~\ref{Eq:sensitivityofobjective1} in association with the chain rule given in Eq.~\ref{Eq:ChainRule}, one can find the sensitivity of $\mathcal{L}$ with respect to the design vector, i.e., $\frac{d\mathcal{L}}{d{\bm{\rho}}}$. This formulation facilitates straightforward evaluation of the load sensitivities (Eq.~\ref{Eq:sensitivityofobjective2}) that affect the optimized designs of Pa-CMs \citep{kumar2020topology} and therefore, are important to consider while designing such mechanisms.
	\section{Numerical examples and discussions}\label{Sec:Numericalexamples}
	
	This section evaluates the presented robust approach by designing pressure-actuated  inverter and gripper CMs.  The symmetric half design domains for designing these mechanisms are displayed in Fig.~\ref{fig:DesignDomain}. $L_x=\SI{0.2}{\meter}$ and $L_y=\SI{0.1}{\meter}$ are set, where $L_x$ and $L_y$ represent the dimension in $x-$ and $y-$directions, respectively. $\SI{1}{\bar}$ pressure load is applied on the left edge of the domains, whereas remaining edges excluding the symmetric ones experience zero pressure loading. The fixed parts of the domains and their symmetry boundaries are also depicted.  Springs with spring stiffnesses $k_\text{ss}=\SI{1e4}{\newton\per\meter}$ represent the workpiece at mechanism output locations (Fig.~\ref{fig:DesignDomain}).  Table~\ref{Table:T1} summarizes the design parameters used in the optimization. We use $N_\text{ex}\times N_\text{ey} = 200 \times 100$ bi-linear quadrilateral FEs to parameterize the symmetric half design domains (Fig.~\ref{fig:DesignDomain}), where $N_\text{ex}$ and $N_\text{ey}$ indicate FEs in $x-$ and $y-$directions,  respectively. One can also employ honeycomb tessellation (hexagonal FEs) \citep{kumar2022honeytop90} for the design representation. A density-based TO approach with one design variable for each FE is employed with plane stress conditions. The design variable is considered constant within each FE. The external move limit of the MMA optimizer is set to $0.1$.  The color schemes displayed in Fig.~\ref{fig:colorschem} are used to plot material, pressure and displacement fields in this paper.
	
	\subsection{Traditional pressure-actuated inverter mechanism}\label{Sec:Traditional Pa-IV-CM}
	The symmetric half design domain displayed in Fig.~\ref{fig:inverter} is considered, and the optimization formulation presented in \citet{kumar2020topology} is employed for designing the inverter mechanism.  The filter radius $r_\text{fill}$ is set to $2.5\times\max\left(\frac{L_x}{N_\text{ex}},\,\frac{L_y}{N_\text{ey}}\right)$. 20$\%$ material volume is permitted. 
	
	Fig.~\ref{fig:Inverterwithoutrobust}a depicts the optimized inverter mechanism with its final pressure field. $M_\text{nd}=8.9\%$ and $\Delta = \SI{0.0235}{\milli\meter}$ are found. Insets in Fig.~\ref{fig:Inverterwithoutrobust}a display the thin flexure regions constituted of gray elements. These geometrical anomalies  pose challenges in manufacturing and thus, they are undesirable.  When the optimized design is approximated using: (i) $\rho\in [0.75,\,1] \to  1$ and $\rho\in [0,\,0.75) \to  0$ as shown in Fig.~\ref{fig:Inverterwithoutrobust}a$_1$,  a design with thin and potentially challenging to manufacture regions is obtained with $\Delta = \SI{0.0243}{\milli\meter}$ and (ii) $\rho\in[0.25,\,1]\to  1$ and  $\rho\in[0,\,0.25)\to  0$ as displayed in Fig.~\ref{fig:Inverterwithoutrobust}a$_2$, a  design with $\Delta = \SI{0.0205}{\milli \meter}$ is obtained, which is considerably lower than the displacement obtained for the actual design (Fig.~\ref{fig:Inverterwithoutrobust}a). These approximations also alter the topologies. Therefore, to circumvent these issues and also, to obtain optimized solutions close to 0-1 such that contours of the designs can be extracted without performing any approximation for fabrication purposes, as mentioned before, the robust formulation is employed in all following examples \citep{wang2011projection}.

	\subsection{Pressure-actuated robust mechanisms}\label{Sec:Pa IV and GP}
	The various optimized designs of the pressure-actuated inverter, gripper and contractor CMs are presented using the robust formulation (Eq.~\ref{Eq:actualoptimization}). In each case, we get three optimized designs, i.e., the dilated, intermediate and eroded continua, and in those, the intermediate designs are intended for fabrication. 
	
	The permitted volume fraction for the intermediate design is set to 0.20 for all the cases. The maximum number of MMA iterations is fixed to 400. In the projection filter (Eq.~\ref{Eq:projectionfilter}), $\beta$ is altered from 1 to $128$ using a continuation scheme wherein $\beta$ is doubled after each $50$ MMA iteration and once it reaches to 128, it remains so for the remaining optimization iterations. The volume update for the dilated design is performed at each 25$^\text{th}$ MMA iteration.

	\begin{figure*}
		\begin{subfigure}[t]{0.45\textwidth}
			\centering
			\begin{tikzpicture}
				\pgfplotsset{compat = 1.3}
				\begin{axis}[
					width = 1\textwidth,
					xlabel=MMA iteration,
					xlabel=MMA iteration,
					ylabel= -$1000\times\frac{MSE}{SE}$,
					legend style={at={(1.200,0.65),font=\fontsize{4}{5}\tiny},anchor=east}]
					\pgfplotstableread{a2do.txt}\mydata;
					\addplot[smooth,black,,mark size=1pt,style={thick}]
					table {\mydata};
					\pgfplotstableread{a2io.txt}\mydata;
					\addplot[smooth,blue,mark size=1pt,style={thick}]
					table {\mydata};
					\pgfplotstableread{a2eo.txt}\mydata;
					\addplot[smooth,red,mark size=1pt,style={thick}]
					table {\mydata};
				\end{axis}
			\end{tikzpicture}
			\caption{}
			\label{fig:IV2Oplot}
		\end{subfigure}
		\quad
		\begin{subfigure}[t]{0.45\textwidth}
			\centering
			\begin{tikzpicture} 	
				\pgfplotsset{compat = 1.3}
				\begin{axis}[
					width = 1\textwidth,
					xlabel=MMA iteration,
					ylabel= -$1000\times\frac{MSE}{SE}$,
					legend style={at={(1.200,0.65),font=\fontsize{4}{5}\tiny},anchor=east}]
					\pgfplotstableread{a3do.txt}\mydata;
					\addplot[smooth,black,,mark size=1pt,style={thick}]
					table {\mydata};
					\pgfplotstableread{a3io.txt}\mydata;
					\addplot[smooth,blue,mark size=1pt,style={thick}]
					table {\mydata};
					\pgfplotstableread{a3eo.txt}\mydata;
					\addplot[smooth,red,mark size=1pt,style={thick}]
					table {\mydata};
				\end{axis}
			\end{tikzpicture}
			\caption{}
			\label{fig:IV3Oplot}
		\end{subfigure}
		\caption{Objective  convergence plots for IV2  and IV3 Pa-CMs are displayed in (\subref{fig:IV2Oplot}) and (\subref{fig:IV3Oplot}) respectively. The black, blue and red curves indicate the dilated, intermediate and eroded designs convergence history. We use the same color scheme for showing convergence curves henceforth.}
		\label{fig:InverterOConvergence}
	\end{figure*}
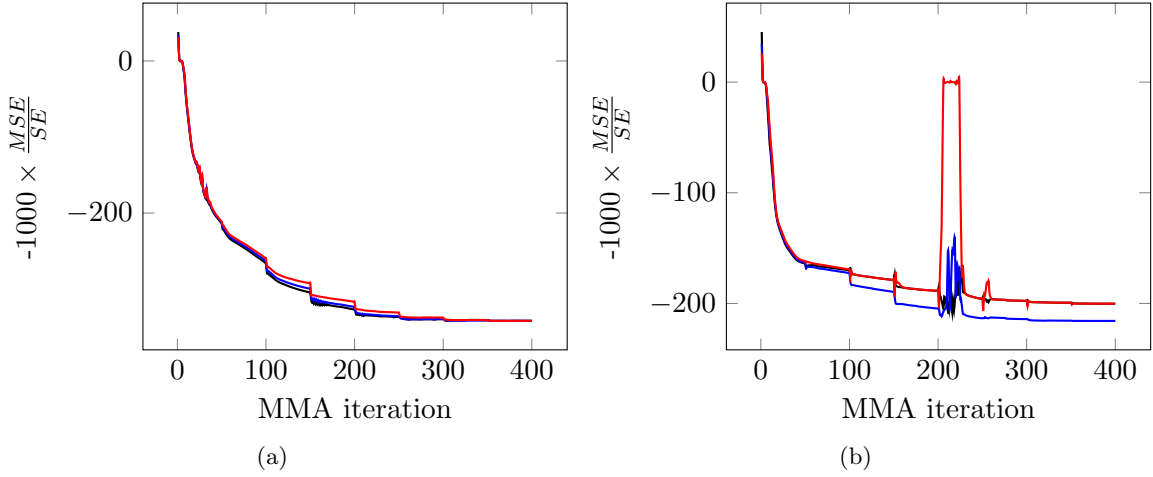
	
	
	\begin{figure*}
		\begin{subfigure}[t]{0.45\textwidth}
			\centering
			\begin{tikzpicture}
				\pgfplotsset{compat = 1.3}
				\begin{axis}[
					width = 1\textwidth,
					xlabel=MMA iteration,
					ylabel= Volume fraction,
					ytick ={0,0.1,0.15,0.20, 0.25}]
					\pgfplotstableread{a2dv.txt}\mydata;
					\addplot[smooth,black,,mark size=1pt, style={thick}]
					table {\mydata};
					\pgfplotstableread{a2iv.txt}\mydata;
					\addplot[smooth,blue,mark size=1pt, style={thick}]
					table {\mydata};
					\pgfplotstableread{a2ev.txt}\mydata;
					\addplot[smooth,red,mark size=1pt,style={thick}]
					table {\mydata};
				\end{axis}
			\end{tikzpicture}
			\caption{}
			\label{fig:IV2Vplot}
		\end{subfigure}
		\quad
		\begin{subfigure}[t]{0.45\textwidth}
			\centering
			\begin{tikzpicture} 	
				\pgfplotsset{compat = 1.3}
				\begin{axis}[
					width = 1\textwidth,
					xlabel=MMA iteration,
					ylabel= Volume fraction,
					ytick ={0,0.1,0.15,0.20, 0.25},
					legend style={at={(1.520,0.45)},font=\fontsize{4}{5}\tiny,anchor=east}]
					\pgfplotstableread{a3dv.txt}\mydata;
					\addplot[smooth,black,,mark size=1pt, style={thick}]
					table {\mydata};
					\pgfplotstableread{a3iv.txt}\mydata;
					\addplot[smooth,blue,mark size=1pt, style={thick}]
					table {\mydata};
					\pgfplotstableread{a3ev.txt}\mydata;
					\addplot[smooth,red,mark size=1pt,style={thick}]
					table {\mydata};
				\end{axis}
			\end{tikzpicture}
			\caption{}
			\label{fig:IV3Vplot}
		\end{subfigure}
		\caption{Volume fraction convergence plots for IV2 and IV3 Pa-CMs are displayed in (\subref{fig:IV2Vplot}) and (\subref{fig:IV3Vplot}) respectively.}
		\label{fig:InverterVConvergence}
	\end{figure*}

	\begin{figure*}
		\begin{subfigure}[t]{0.45\textwidth}
			\centering
			\begin{tikzpicture}
				\pgfplotsset{compat = 1.3}
				\begin{axis}[
					width = 1\textwidth,
					xlabel=MMA iteration,
					xlabel=MMA iteration,
					ylabel= -$1000\times\frac{MSE}{SE}$,
					legend style={at={(1.200,0.65),font=\fontsize{4}{5}\tiny},anchor=east}]
					\pgfplotstableread{b2do.txt}\mydata;
					\addplot[smooth,black,,mark size=1pt,style={thick}]
					table {\mydata};
					\pgfplotstableread{b2io.txt}\mydata;
					\addplot[smooth,blue,mark size=1pt,style={thick}]
					table {\mydata};
					\pgfplotstableread{b2eo.txt}\mydata;
					\addplot[smooth,red,mark size=1pt,style={thick}]
					table {\mydata};
				\end{axis}
			\end{tikzpicture}
			\caption{}
			\label{fig:GP2Oplot}
		\end{subfigure}
		\quad
		\begin{subfigure}[t]{0.45\textwidth}
			\centering
			\begin{tikzpicture} 	
				\pgfplotsset{compat = 1.3}
				\begin{axis}[
					width = 1\textwidth,
					xlabel=MMA iteration,
					ylabel= -$1000\times\frac{MSE}{SE}$,
					legend style={at={(1.200,0.65),font=\fontsize{4}{5}\tiny},anchor=east}]
					\pgfplotstableread{b3do.txt}\mydata;
					\addplot[smooth,black,,mark size=1pt,style={thick}]
					table {\mydata};
					\pgfplotstableread{b3io.txt}\mydata;
					\addplot[smooth,blue,mark size=1pt,style={thick}]
					table {\mydata};
					\pgfplotstableread{b3eo.txt}\mydata;
					\addplot[smooth,red,mark size=1pt,style={thick}]
					table {\mydata};
				\end{axis}
			\end{tikzpicture}
			\caption{}
			\label{fig:GP3Oplot}
		\end{subfigure}
		\caption{Objective  convergence plots for GP2 and GP3 Pa-CMs are shown in (\subref{fig:GP2Oplot}) and (\subref{fig:GP3Oplot}) respectively.}
		\label{fig:GripperOConvergence}
	\end{figure*}
	
	\begin{figure*}
		\begin{subfigure}[t]{0.45\textwidth}
			\centering
			\begin{tikzpicture}
				\pgfplotsset{compat = 1.3}
				\begin{axis}[
					width = 1\textwidth,
					xlabel=MMA iteration,
					ylabel= Volume fraction,
					ytick ={0,0.1,0.15,0.20, 0.25}]
					\pgfplotstableread{b2dv.txt}\mydata;
					\addplot[smooth,black,,mark size=1pt, style={thick}]
					table {\mydata};
					\pgfplotstableread{b2iv.txt}\mydata;
					\addplot[smooth,blue,mark size=1pt, style={thick}]
					table {\mydata};
					\pgfplotstableread{b2ev.txt}\mydata;
					\addplot[smooth,red,mark size=1pt,style={thick}]
					table {\mydata};
				\end{axis}
			\end{tikzpicture}
			\caption{}
			\label{fig:GP2Vplot}
		\end{subfigure}
		\quad
		\begin{subfigure}[t]{0.45\textwidth}
			\centering
			\begin{tikzpicture} 	
				\pgfplotsset{compat = 1.3}
				\begin{axis}[
					width = 1\textwidth,
					xlabel=MMA iteration,
					ylabel= Volume fraction,
					ytick ={0,0.1,0.15,0.20, 0.25},
					legend style={at={(1.520,0.45)},font=\fontsize{4}{5}\tiny,anchor=east}]
					\pgfplotstableread{b3dv.txt}\mydata;
					\addplot[smooth,black,,mark size=1pt, style={thick}]
					table {\mydata};
					\pgfplotstableread{b3iv.txt}\mydata;
					\addplot[smooth,blue,mark size=1pt, style={thick}]
					table {\mydata};
					\pgfplotstableread{b3ev.txt}\mydata;
					\addplot[smooth,red,mark size=1pt,style={thick}]
					table {\mydata};
				\end{axis}
			\end{tikzpicture}
			\caption{}
			\label{fig:GP3Vplot}
		\end{subfigure}
		\caption{Volume fraction convergence plots for GP2 and GP3 Pa-CMs are shown in (\subref{fig:GP2Vplot}) and (\subref{fig:GP3Vplot}) respectively.}
		\label{fig:GripperVConvergence}
	\end{figure*}

	\begin{figure*}
		\begin{subfigure}[t]{0.45\textwidth}
			\centering
			\begin{tikzpicture}
				\pgfplotsset{compat = 1.3}
				\begin{axis}[
					width = 1\textwidth,
					xlabel=MMA iteration,
					xlabel=MMA iteration,
					ylabel= -$1000\times\frac{MSE}{SE}$,
					legend style={at={(1.200,0.65),font=\fontsize{4}{5}\tiny},anchor=east}]
					\pgfplotstableread{c2do.txt}\mydata;
					\addplot[smooth,black,,mark size=1pt,style={thick}]
					table {\mydata};
					\pgfplotstableread{c2io.txt}\mydata;
					\addplot[smooth,blue,mark size=1pt,style={thick}]
					table {\mydata};
					\pgfplotstableread{c2eo.txt}\mydata;
					\addplot[smooth,red,mark size=1pt,style={thick}]
					table {\mydata};
				\end{axis}
			\end{tikzpicture}
			\caption{}
			\label{fig:CT2Oplot}
		\end{subfigure}
		\quad
		\begin{subfigure}[t]{0.45\textwidth}
			\centering
			\begin{tikzpicture} 	
				\pgfplotsset{compat = 1.3}
				\begin{axis}[
					width = 1\textwidth,
					xlabel=MMA iteration,
					ylabel= -$1000\times\frac{MSE}{SE}$,
					legend style={at={(1.200,0.65),font=\fontsize{4}{5}\tiny},anchor=east}]
					\pgfplotstableread{c3do.txt}\mydata;
					\addplot[smooth,black,,mark size=1pt,style={thick}]
					table {\mydata};
					\pgfplotstableread{c3io.txt}\mydata;
					\addplot[smooth,blue,mark size=1pt,style={thick}]
					table {\mydata};
					\pgfplotstableread{c3eo.txt}\mydata;
					\addplot[smooth,red,mark size=1pt,style={thick}]
					table {\mydata};
				\end{axis}
			\end{tikzpicture}
			\caption{}
			\label{fig:CT3Oplot}
		\end{subfigure}
		\caption{Objective convergence plots for CT2 and CT3 Pa-CMs are displayed in (\subref{fig:CT2Oplot}) and (\subref{fig:CT3Oplot}) respectively.}
		\label{fig:ContractorOConvergence}
	\end{figure*}
	
	\begin{figure*}
		\begin{subfigure}[t]{0.45\textwidth}
			\centering
			\begin{tikzpicture}
				\pgfplotsset{compat = 1.3}
				\begin{axis}[
					width = 1\textwidth,
					xlabel=MMA iteration,
					ylabel= Volume fraction,
					ytick ={0,0.1,0.15,0.20, 0.25}]
					\pgfplotstableread{c2dv.txt}\mydata;
					\addplot[smooth,black,,mark size=1pt, style={thick}]
					table {\mydata};
					\pgfplotstableread{c2iv.txt}\mydata;
					\addplot[smooth,blue,mark size=1pt, style={thick}]
					table {\mydata};
					\pgfplotstableread{c2ev.txt}\mydata;
					\addplot[smooth,red,mark size=1pt,style={thick}]
					table {\mydata};
				\end{axis}
			\end{tikzpicture}
			\caption{}
			\label{fig:CT2Vplot}
		\end{subfigure}
		\quad
		\begin{subfigure}[t]{0.45\textwidth}
			\centering
			\begin{tikzpicture} 	
				\pgfplotsset{compat = 1.3}
				\begin{axis}[
					width = 1\textwidth,
					xlabel=MMA iteration,
					ylabel= Volume fraction,
					ytick ={0,0.1,0.15,0.20, 0.25},
					legend style={at={(1.520,0.45)},font=\fontsize{4}{5}\tiny,anchor=east}]
					\pgfplotstableread{c3dv.txt}\mydata;
					\addplot[smooth,black,,mark size=1pt, style={thick}]
					table {\mydata};
					\pgfplotstableread{c3iv.txt}\mydata;
					\addplot[smooth,blue,mark size=1pt, style={thick}]
					table {\mydata};
					\pgfplotstableread{c3ev.txt}\mydata;
					\addplot[smooth,red,mark size=1pt,style={thick}]
					table {\mydata};
				\end{axis}
			\end{tikzpicture}
			\caption{}
			\label{fig:CT3Vplot}
		\end{subfigure}
		\caption{Volume fraction convergence plots for CT2 and CT3 Pa-CMs are displayed in (\subref{fig:CT2Vplot}) and (\subref{fig:CT3Vplot}) respectively.}
		\label{fig:ContractorVConvergence}
	\end{figure*}

	\begin{figure*}
		\begin{subfigure}[t]{0.30\textwidth}
			\centering
			\includegraphics[scale = 0.35]{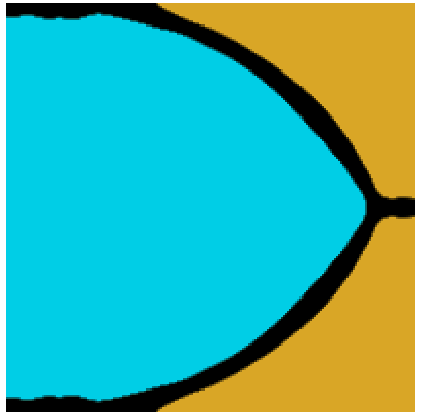}
			\caption{$\Delta = \SI{-0.537}{\milli \meter}$}
			$-\frac{MSE}{SE}= -0.14,\,SE = \SI{3.94e-3}{\newton \meter}$
			\label{fig:IV3_V01i}
		\end{subfigure}
		\begin{subfigure}[t]{0.30\textwidth}
			\centering
			\includegraphics[scale =0.35]{a3i}
			\caption{$\Delta = \SI{-0.29}{\milli \meter}$}
			$-\frac{MSE}{SE}= -0.22,\,SE = \SI{1.32e-3}{\newton \meter}$	
			\label{fig:a3i}
		\end{subfigure}
		\begin{subfigure}[t]{0.30\textwidth}
			\centering
			\includegraphics[scale =0.35]{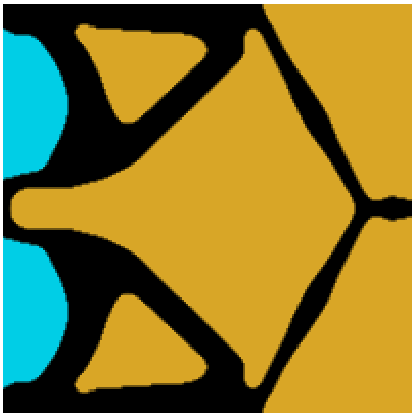}
			\caption{$\Delta = \SI{-0.19}{\milli \meter}$}
			$-\frac{MSE}{SE}= -0.38,\,SE = \SI{5.01e-4}{\newton \meter}$
			\label{fig:IV3_V03i}
		\end{subfigure}
		\begin{subfigure}[t]{0.30\textwidth}
			\centering
			\includegraphics[scale = 0.35]{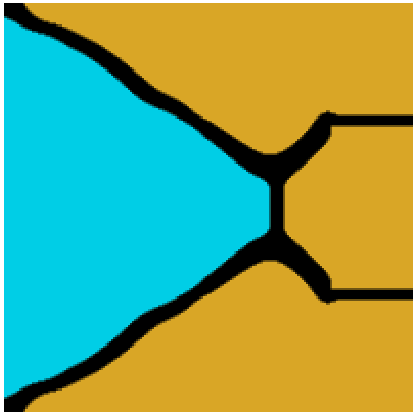}
			\caption{$\Delta = \SI{-0.141}{\milli \meter}$}
			\hspace{-2em}	$-\frac{MSE}{SE}= -0.015,\,SE = \SI{9.37e-3}{\newton \meter}$
			\label{fig:GP3_V01i}
		\end{subfigure}
		\begin{subfigure}[t]{0.30\textwidth}
			\centering
			\includegraphics[scale = 0.35]{b3i}
			\caption{$\Delta = \SI{-0.189}{\milli \meter}$}
			$-\frac{MSE}{SE}= -0.29,\,SE = \SI{6.62e-4}{\newton \meter}$
			\label{fig:b3i}
		\end{subfigure}
		\begin{subfigure}[t]{0.30\textwidth}
			\centering
			\includegraphics[scale = 0.35]{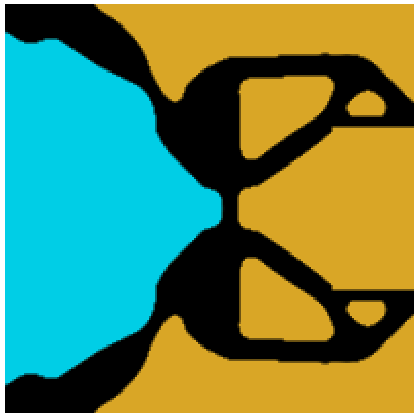}
			\caption{$\Delta = \SI{-0.167}{\milli \meter}$}
			$-\frac{MSE}{SE}= -0.34,\,SE = \SI{4.88e-4}{\newton \meter}$
			\label{fig:GP3_V03i}
		\end{subfigure}
		\begin{subfigure}[t]{0.30\textwidth}
			\centering
			\includegraphics[scale = 0.35]{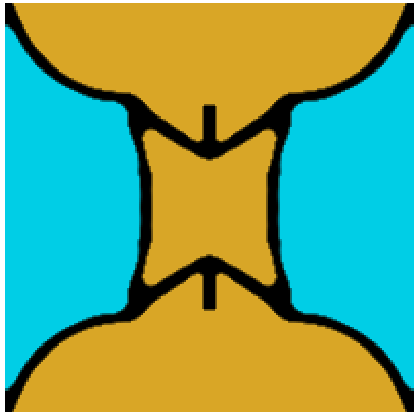}
			\caption{$\Delta = \SI{-0.0958}{\milli \meter}$}
			$-\frac{MSE}{SE}= -0.13,\,SE = \SI{7.73e-4}{\newton \meter}$
			\label{fig:CT3_V01i}
		\end{subfigure}
		\begin{subfigure}[t]{0.30\textwidth}
			\centering
			\includegraphics[scale = 0.35]{c3i}
			\caption{$\Delta = \SI{-0.041}{\milli \meter}$}
			$-\frac{MSE}{SE}= -0.23,\,SE = \SI{1.81e-4}{\newton \meter}$
			\label{fig:c3i}
		\end{subfigure}
		\begin{subfigure}[t]{0.30\textwidth}
			\centering
			\includegraphics[scale = 0.35]{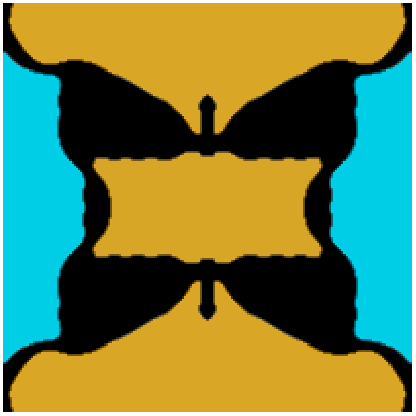}
			\caption{$\Delta = \SI{-0.0363}{\milli \meter}$}
			$-\frac{MSE}{SE}= -0.30,\,SE = \SI{1.23e-4}{\newton \meter}$
			\label{fig:CT3_V03i}
		\end{subfigure}
		\caption{Intermediate optimized designs of inverter (Row~1), gripper (Row~2) and contractor (Row~3) Pa-CMs. Results in column~1, column~2 and column~3 are obtained using $V^*_f = 0.1$, $V^*_f = 0.20$ and $V^*_f=0.3$ respectively.}	\label{fig:Volumefractionsresults}
	\end{figure*}

	\begin{figure*}
		\begin{subfigure}[t]{0.18\textwidth}
			\centering
			\includegraphics[scale = 0.2]{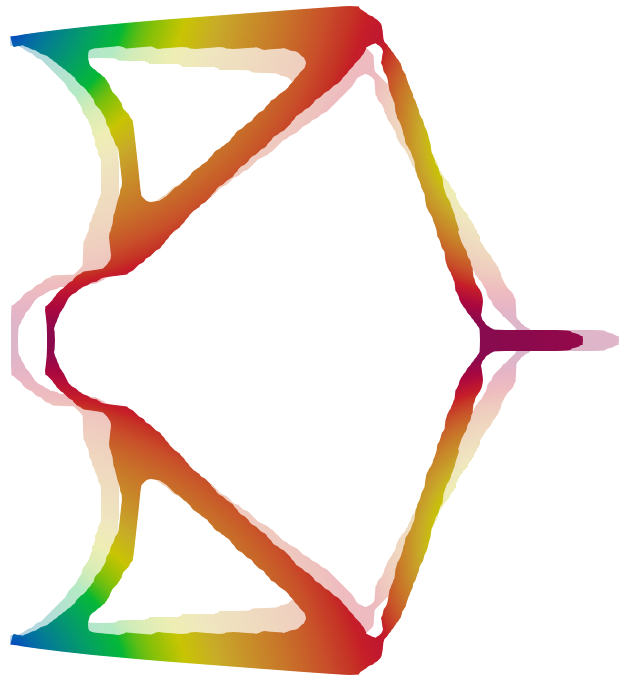}
			\caption{IV2}
			\label{fig:IV2L_50}
		\end{subfigure}
		\begin{subfigure}[t]{0.18\textwidth}
			\centering
			\includegraphics[scale = 0.2]{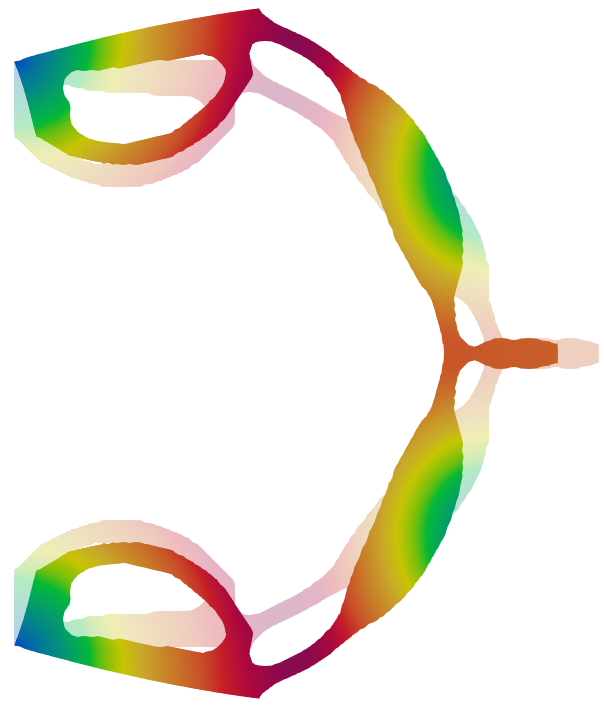}
			\caption{IV3}
			\label{fig:IV3L_50}
		\end{subfigure}
		\begin{subfigure}[t]{0.18\textwidth}
			\centering
			\includegraphics[scale = 0.21]{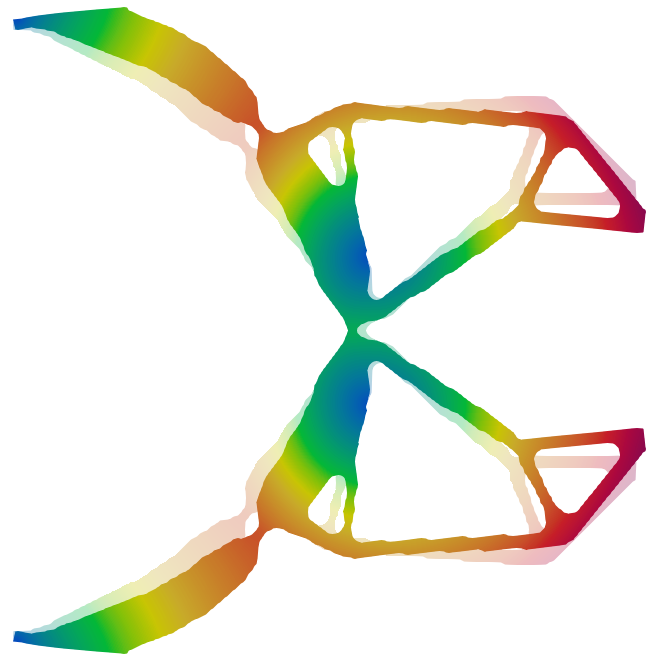}
			\caption{GP2}
			\label{fig:GP2L_50}
		\end{subfigure}
		\quad
		\begin{subfigure}[t]{0.18\textwidth}
			\centering
			\includegraphics[scale = 0.2]{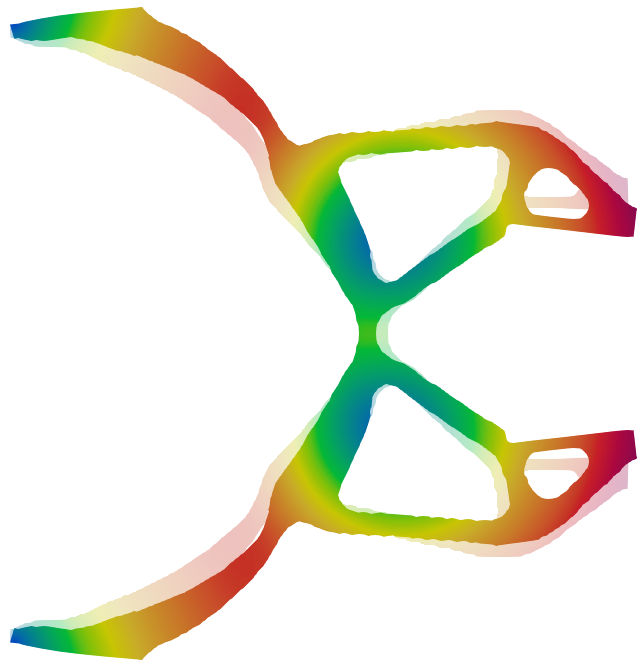}
			\caption{GP3}
			\label{fig:GP3L_50}
		\end{subfigure}
		\quad
		\begin{subfigure}[t]{0.18\textwidth}
			\centering
			\includegraphics[scale = 0.2]{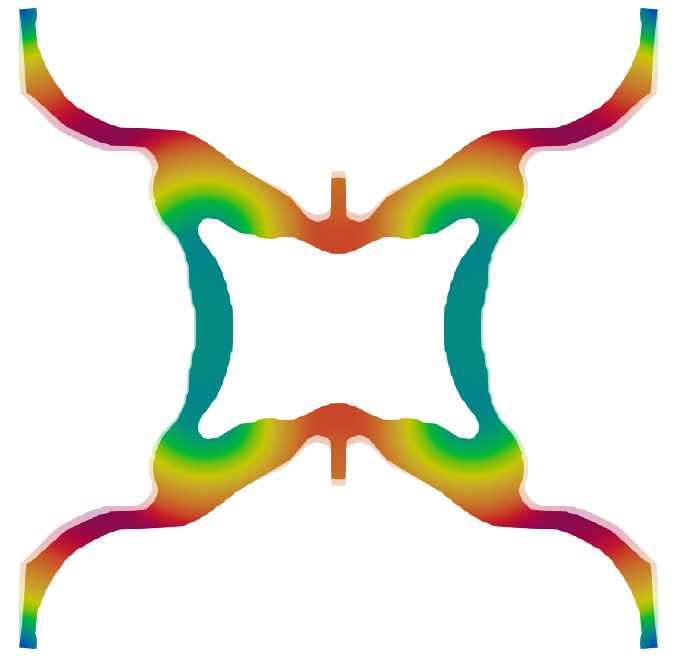}
			\caption{CT3}
			\label{fig:CT3L_50}
		\end{subfigure}
		\caption{Deformed profiles of the optimized inverter, gripper and contractor CMs are displayed with 50 times magnified displacement obtained from linear analysis. }	\label{fig:deformedprofiles}
	\end{figure*}

	In case of the inverter CMs, it is desired that the mechanisms provide deformation in the opposite direction of the pressure loading direction, whereas gripping motions are desired in response to the pressure load for the gripper CMs.  At the output location of the gripper mechanism,  a void area of size $\frac{L_x}{5}\times \frac{L_x}{5}$ and  a solid region of area $\frac{L_x}{5}\times \frac{L_x}{40}$ are considered to facilitate gripping of a workpiece. For the contracting mechanism, a contraction motion is sought while applying pressure loads on the left and right edges. A solid non-design domain of size $\frac{L_x}{40}\times \frac{L_y}{4}$ is present in the middle of the symmetric half domain of the contractor mechanism. Different filter radii and $\Delta\eta$ are used to find the mechanism optimized designs.
	
	The full final inverter, gripper  and contractor mechanisms are obtained by suitably transforming the symmetric half optimized results, and they are depicted in Fig.~\ref{fig:robustinverters}, Fig.~\ref{fig:robustgripper}, and Fig.~\ref{fig:robustcontractor} respectively. Note the absence of one-node hinges--these would lead to disconnected structures in the eroded design with very poor performance, and hence the optimizer avoids such problematic features entirely. The topology for the dilated, intermediate and eroded is the same for all the presented cases. The eroded designs feature thin members, whereas the dilated designs consist of  thicker branches, which is expected. Thicknesses of the members of the intermediate designs are between those of the respective eroded and dilated designs. $M_\text{nd}$ and final volume fraction for each optimized design are also mentioned. The obtained solutions in Fig.~\ref{fig:robustinverters}, Fig.~\ref{fig:robustgripper} and Fig.~\ref{fig:robustcontractor}are very close to 0-1 designs as their $M_\text{nd}$ values are low. This is a very desirable result, since now the design interpretation no longer significantly changes the geometry of the design and its performance. 
	
	One can note that IV2 Pa-CMs have different topologies than those of IV1, IV3 and IV4 Pa-CMs (Fig.~\ref{fig:robustinverters}). Fig.~\ref{fig:IV1_MMA} and Fig.~\ref{fig:IV2_MMA} display the topologies of IV1 Pa-CMs and IV2 Pa-CMs at different MMA iterations, respectively. Only the degree of manufacturing error, expressed by parameter $\Delta\eta$, differs between these two cases. From Figs.~\ref{fig:IV1_5MMA}-\ref{fig:IV1_6MMA} and Figs.~\ref{fig:IV2_5MMA}-\ref{fig:IV2_6MMA}, until 6 MMA iterations, design evolution of IV1 and IV2 mechanisms is similar. At the $7^\text{th}$ MMA iteration, a minor difference can be noted in the material layouts of these mechanisms (Figs.~\ref{fig:IV1_7MMA} and \ref{fig:IV2_7MMA}) and that eventually has lead to entirely different outcomes.  Parameter dependence is expected given the non-convex nature of topology optimization problems. Moreover, IV2 represents the case with the smallest manufacturing error, thus has the highest design freedom and allows for the thinnest features. This apparently allows the optimization process to pursue a different solution with more localized (but not one-node connected) hinges. From a performance viewpoint, design IV1 is superior, hence in case IV2 an inferior local optimum is obtained. This may be the downside of the larger design freedom in this case, as it also can lead to a larger number of local optima. Note that the found output stroke $\Delta$ of inverter mechanisms obtained using the proposed formulation (Fig.~\ref{fig:robustinverters}) is better than that of the inverter mechanism (Sec.~\ref{Sec:Traditional Pa-IV-CM}) obtained using the previous method presented in \citet{kumar2020topology}. This is because the current formulation tends to give relatively better distributed compliance with crisp boundaries, and as per \citet{yin2001topology} such mechanisms perform relatively better.  Because of the improved boundary definition, after post-processing the difference in performance is likely to increase even further. Design extraction is discussed in Sec.~\ref{Sec:ExtOptDesigns}.

	The convergence of the objective function is shown in Fig.~\ref{fig:InverterOConvergence}, Fig.~\ref{fig:GripperOConvergence} and Fig.~\ref{fig:ContractorOConvergence}, and the corresponding evolution of the volume fraction in Fig.~\ref{fig:InverterVConvergence}, Fig.~\ref{fig:GripperVConvergence} and Fig.~\ref{fig:ContractorVConvergence}. Stepwise changes are associated to updates of the parameter $\beta$. In all cases, the final volume fractions of the optimized intermediate designs are observed to be equal to the permitted volume fraction, i.e., 0.20. Near the 200th MMA iteration the objective values corresponding to the eroded designs IV3 Pa-CMs (red curve in \ref{fig:IV3Oplot}) are close to zero, which indicates the instantaneous disconnectednesses in those eroded designs. It is noticed that the optimized mechanisms obtained with $\Delta\eta=0.15$ have larger minimum length scale than those obtained with $\Delta\eta=0.05$ at the same filter radius. In addition, the optimized mechanisms with same $\Delta\eta$ but higher filter radius have larger minimum length scale. Therefore, the minimum feature size increases with increase in $\Delta\eta$ and is also a function of the filter radius, which are known properties of the robust formulation \citep{wang2011projection,trillet2021analytical}. 
	
	The layout of the optimized mechanisms is different than the traditional counterparts except that of IV2 Pa-CMs. The obtained designs contain a large space for fluid to inflate. This is reminiscent of designs of pneumatically-actuated soft robots. The obtained output deformations of the optimized eroded mechanisms are higher than the intermediate designs in each case (Fig.~\ref{fig:robustinverters}, Fig.~\ref{fig:robustgripper} and Fig.~\ref{fig:robustcontractor}), however, as mentioned before, such designs have lower manufacturing limits and as-fabricated designs may be fragile. Figures~\ref{fig:InverterOConvergence} and  \ref{fig:InverterVConvergence} indicate the convergence history of the objective and volume constraints for the optimized IV2 and IV3  mechanisms (Fig.~\ref{fig:robustinverters}). Those for the GP2 and GP3 mechanisms (Fig.~\ref{fig:robustgripper}) are displayed in Fig.~\ref{fig:GripperOConvergence} and Fig.~\ref{fig:GripperVConvergence} respectively. The objective and volume convergence plots for CT2 and CT3 mechanisms are displayed in Fig.~\ref{fig:ContractorOConvergence} and Fig.~\ref{fig:ContractorVConvergence}, respectively. One notices a smooth convergence, and the volume constraint is active for the intermediate design at the end of the optimization for the mechanisms. The objective values  of the intermediate designs are lower than those of the corresponding eroded and dilated designs (Figs.~\ref{fig:InverterOConvergence}, \ref{fig:GripperOConvergence} and \ref{fig:ContractorOConvergence}), indicating that intermediate designs are the best performing ones in the view of the considered multi-criteria objective that determines a balance between the output displacement and strength of the mechanism. The objectives of respective eroded and dilated designs are close to each other at the end of the optimization (Figs.~\ref{fig:InverterOConvergence}, \ref{fig:GripperOConvergence} and \ref{fig:ContractorOConvergence}). The convergence curves of all cases show similar characteristics and relatively orderly behavior (aside from the exptected continuation-induced steps). Showing these characteristics is the main point. The deformed profiles for the full intermediate inverter (IV2 and IV3), gripper (GP2 and GP3) and contractor (CT3) mechanisms at 50 times magnified linear deformation are displayed in Fig.~\ref{fig:deformedprofiles}. We select \{IV2, IV3\} and \{GP2, GP3\} mechanisms to study noting the dissimilarities in their topologies (Figs.~\ref{fig:robustinverters} and \ref{fig:robustgripper}), whereas CT3 mechanism is selected (randomly) for contractor mechanism analyses. The same set of mechanisms is considered in Sec.~\ref{Sec:Largedeformationanalyses} for large deformation analyses with high pressure loads. These Pa-CMs perform as expected, however the deformation profiles are far from those obtained when using nonlinear mechanics with high pressure loads, as studied in Sec.~\ref{Sec:Largedeformationanalyses}. 
	
	\subsection{Pa-CMs for different volume fractions}
	In this section, we demonstrate effects of different volume fractions on the optimized Pa-CMs. $\Delta \eta = 0.15$, filter radius $r_\text{fill} =8.4h$ with $h = \min\left(\frac{L_x}{N_\text{ex}}, \frac{L_y}{N_\text{ey}}\right)$ and $V_f^* = 0.1\,\text{and}\,0.3$ are taken herein.
	
	Figure~\ref{fig:Volumefractionsresults} depicts the intermediate optimized designs of inverter (Row~1), gripper (Row~2) and contractor (Row~3) Pa-CMs. One can note that topologies of the optimized designs for inverter are different with $V^*_f = 0.1$, $V^*_f = 0.2$, and $V^*_f = 0.3$, however those of the contractor mechanisms are same. For the gripper mechanisms, the topologies are different with  $V^*_f = 0.1$, and $V^*_f = 0.3$. By and large, the topologies of optimized CMs depend upon the given volume as different amounts of available material enable different optimal design solutions. In addition, with the lower volume fraction e.g. $V_f^* = 0.1$, the optimized designs contain relatively more area   for fluidic pressure load to inflate like soft robots. Therefore, the final topologies as well as the regions within the optimized CMs to inflate under fluidic pressure loads depend upon the permitted resource volume. One notices that the objective improves as the volume fraction of the mechanisms is increased. This implies that the optimized mechanisms converge towards better combinations of output displacement and stiffness, according to the stated objective (Eq.~\ref{Eq:actualoptimization}). With the increase in volume fraction, strain energy of the mechanism decreases, i.e. stiffness of the mechanisms increase. The output deformation  $\Delta$ decreases as the volume fraction and stiffness increases of the Pa-CM. $\Delta$ can also potentially depend upon the final topology of the mechanism, as noted in gripper Pa-CMs (row 2 of Fig.~\ref{fig:Volumefractionsresults}). It can be concluded that in the current formulation, the volume fraction is an important parameter to explore in Pa-CM design studies.

	\begin{figure*}
		\begin{subfigure}[t]{0.320\textwidth}
			\centering
			\includegraphics[scale = 0.365]{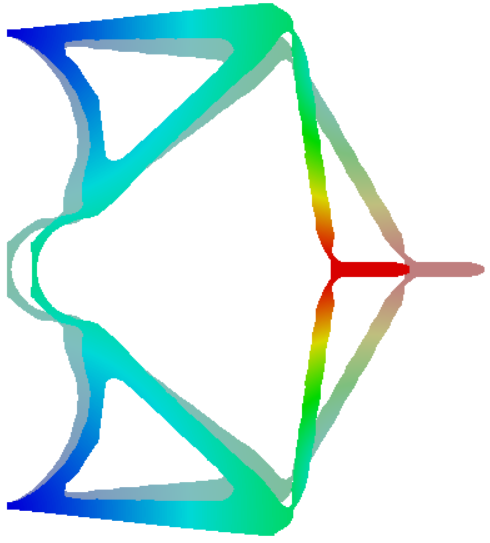}
			\caption{}
			\label{fig:IV2_10}
		\end{subfigure}
		\begin{subfigure}[t]{0.320\textwidth}
			\centering
			\includegraphics[scale = 0.275]{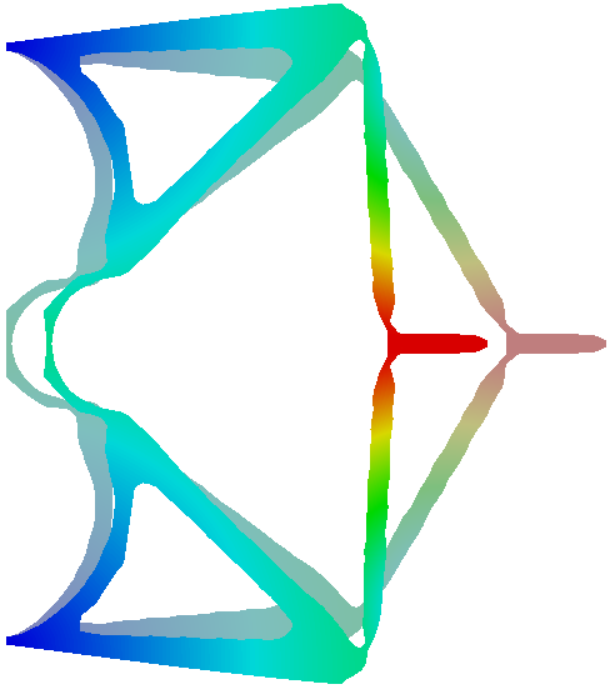}
			\caption{}
			\label{fig:IV2_25}
		\end{subfigure}
		\begin{subfigure}[t]{0.320\textwidth}
			\centering
			\includegraphics[scale = 0.3200]{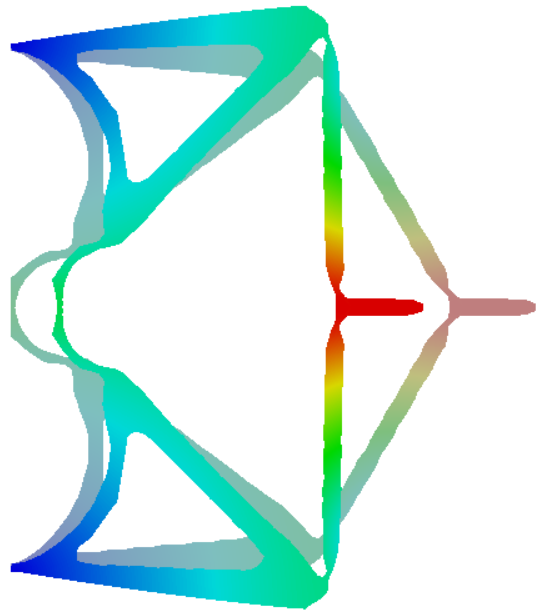}
			\caption{}
			\label{fig:IV2_50}
		\end{subfigure}
		\begin{subfigure}[t]{0.320\textwidth}
			\centering
			\includegraphics[scale = 0.250]{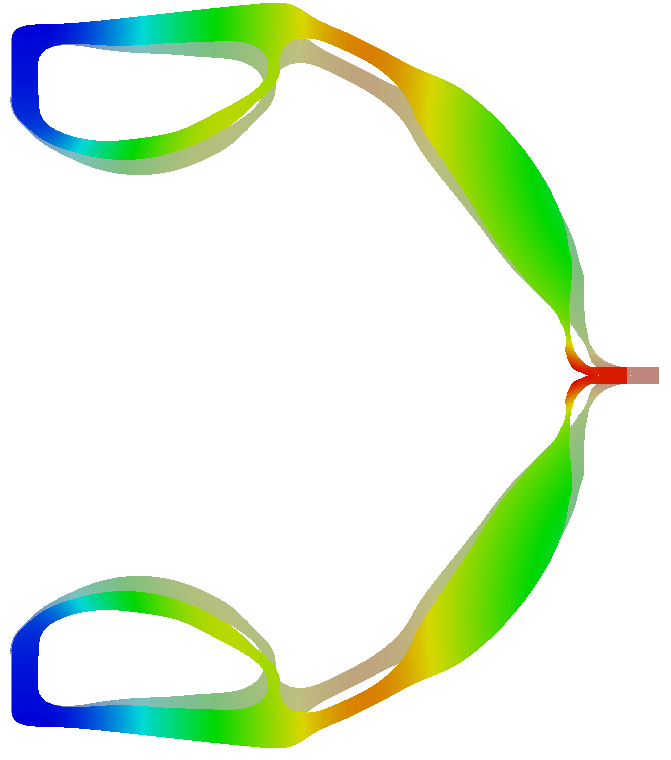}
			\caption{}
			\label{fig:IV3_10}
		\end{subfigure}
		\begin{subfigure}[t]{0.320\textwidth}
			\centering
			\includegraphics[scale = 0.250]{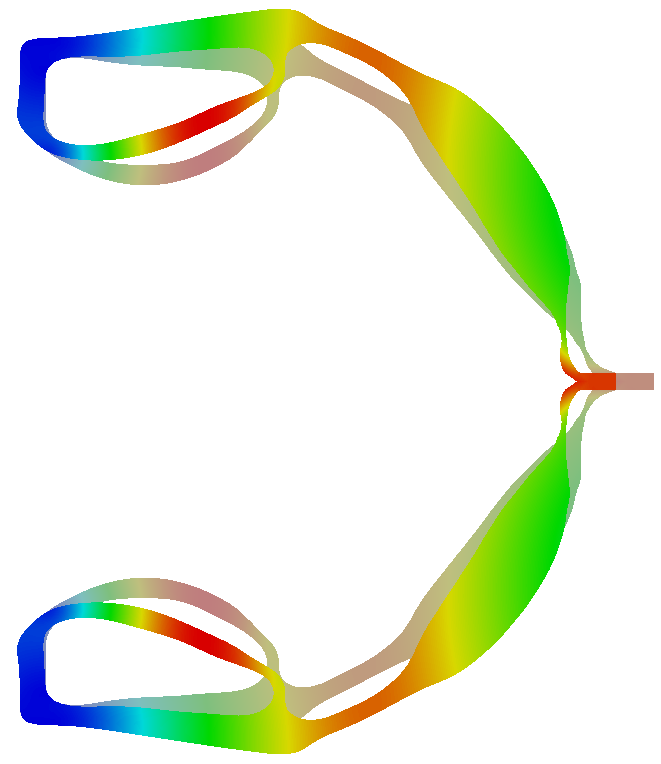}
			\caption{}
			\label{fig:IV3_25}
		\end{subfigure}
		\begin{subfigure}[t]{0.320\textwidth}
			\centering
			\includegraphics[scale = 0.250]{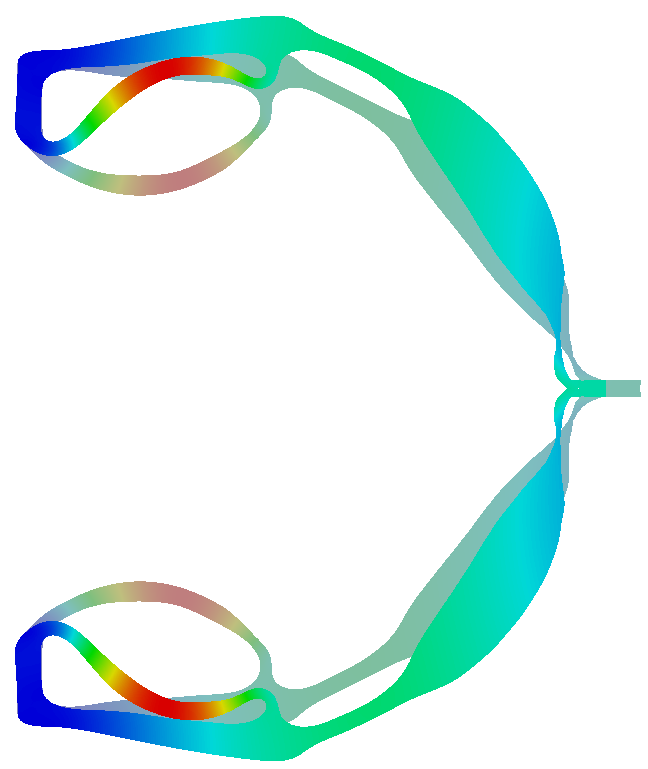}
			\caption{}
			\label{fig:IV3_50}
		\end{subfigure}
		\caption{Deformed profiles of the IV2 and IV3 Pa-CMs at different pressure loads. IV2 Pa-CMs (\subref{fig:IV2_10}) $\SI{10}{\bar}$, (\subref{fig:IV2_25}) $\SI{25}{\bar}$, (\subref{fig:IV2_50}) $\SI{50}{\bar}$ and IV3 Pa-CMs (\subref{fig:IV3_10}) $\SI{10}{\bar}$, (\subref{fig:IV3_25}) $\SI{25}{\bar}$, (\subref{fig:IV3_50}) $\SI{50}{\bar}$. At $\SI{50}{\bar}$ pressure loading, branches of the IV3 Pa-CM come in contact with each other, i.e., \textit{self-contact mode}. Blue and red color indicate minimum and maximum deformation locations, respectively.}	\label{fig:InverterPressure}
	\end{figure*}
	\begin{figure*}
		\begin{subfigure}[t]{0.32\textwidth}
			\centering
			\includegraphics[scale = 0.350]{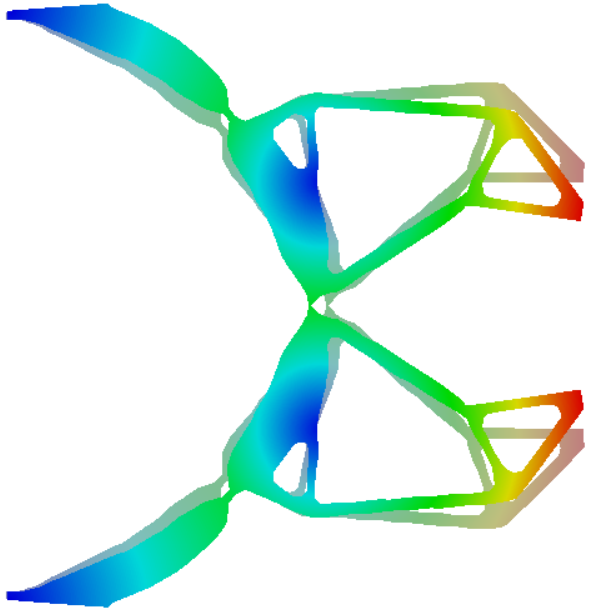}
			\caption{}
			\label{fig:GP2_10}
		\end{subfigure}
		\begin{subfigure}[t]{0.32\textwidth}
			\centering
			\includegraphics[scale = 0.350]{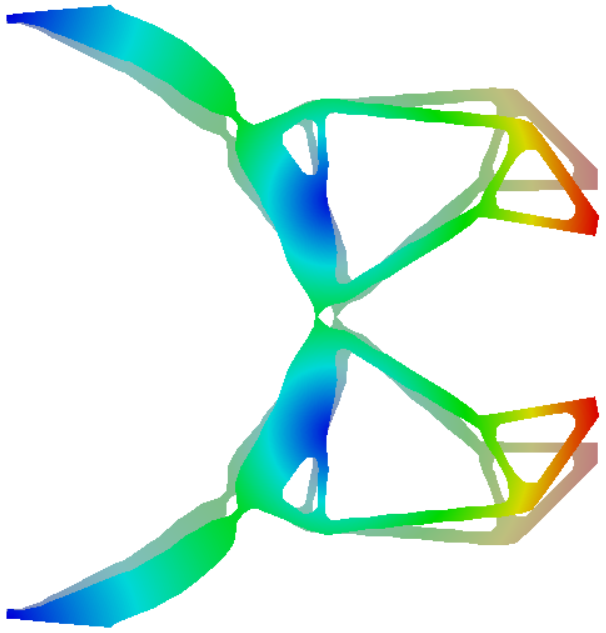}
			\caption{}
			\label{fig:GP2_25}
		\end{subfigure}
		\begin{subfigure}[t]{0.32\textwidth}
			\centering
			\includegraphics[scale = 0.350]{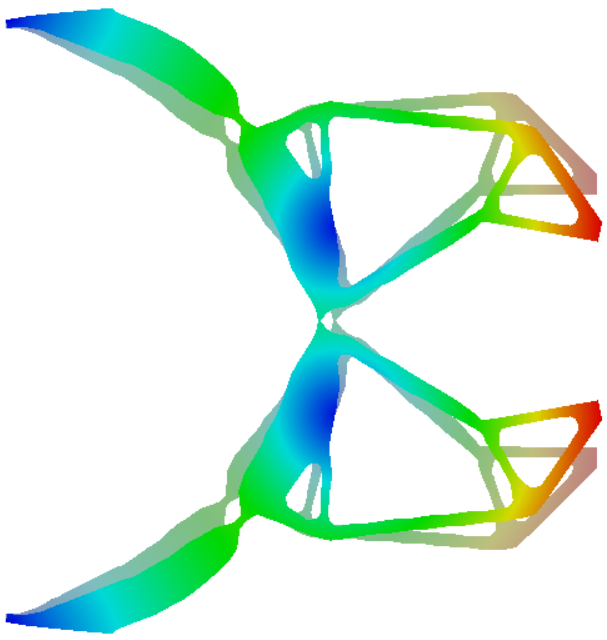}
			\caption{}
			\label{fig:GP2_50}
		\end{subfigure}
		\begin{subfigure}[t]{0.32\textwidth}
			\centering
			\includegraphics[scale = 0.350]{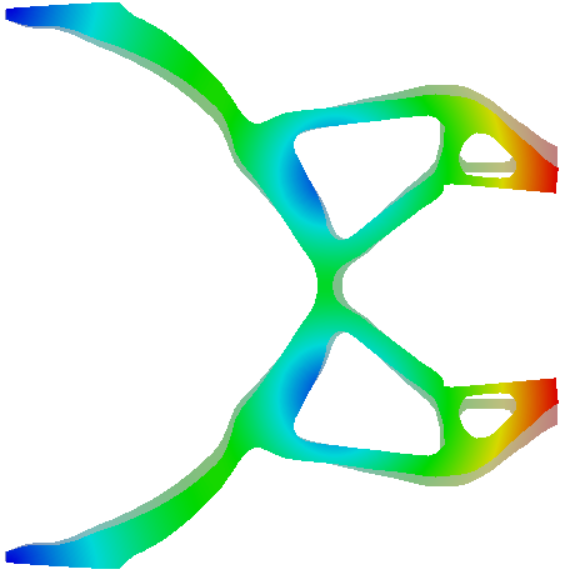}
			\caption{}
			\label{fig:GP3_10}
		\end{subfigure}
		\begin{subfigure}[t]{0.32\textwidth}
			\centering
			\includegraphics[scale = 0.350]{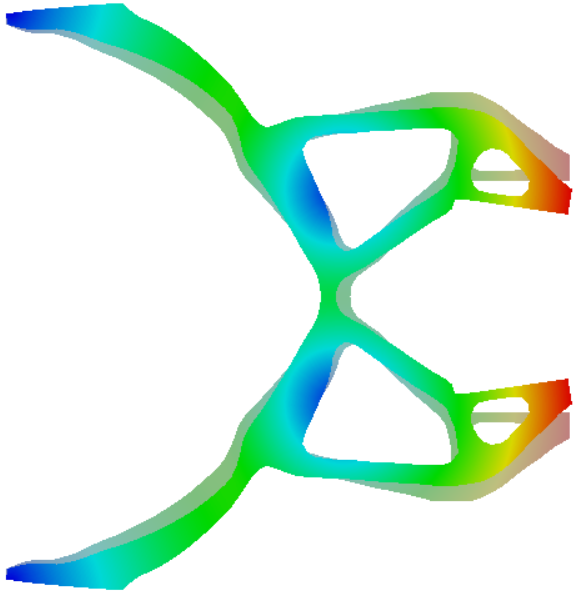}
			\caption{}
			\label{fig:GP3_25}
		\end{subfigure}
		\begin{subfigure}[t]{0.32\textwidth}
			\centering
			\includegraphics[scale = 0.350]{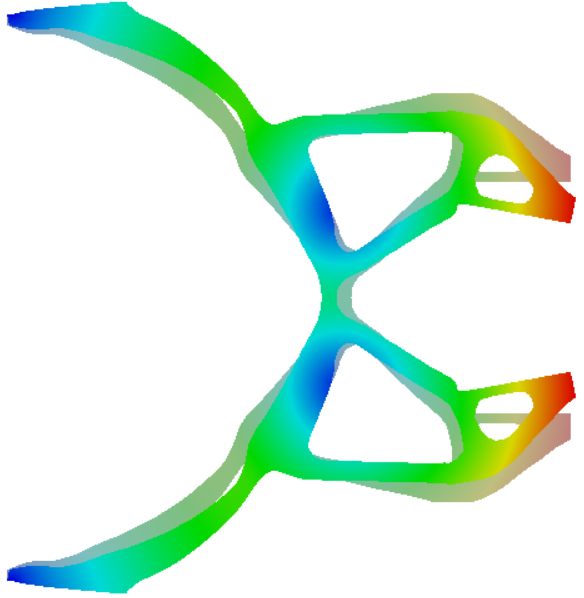}
			\caption{}
			\label{fig:GP3_50}
		\end{subfigure}
		\caption{Deformed profiles of the gripper CM at different pressure loads. GP2 Pa-CMs (\subref{fig:GP2_10}) $\SI{10}{\bar}$, (\subref{fig:GP2_25}) $\SI{25}{\bar}$, (\subref{fig:GP2_50}) $\SI{50}{\bar}$ and GP3 Pa-CMs (\subref{fig:GP3_10}) $\SI{10}{\bar}$, (\subref{fig:GP3_25}) $\SI{25}{\bar}$, (\subref{fig:GP3_50}) $\SI{50}{\bar}$.}	\label{fig:GripperPressure}
	\end{figure*}

	\begin{figure*}
		\begin{subfigure}[t]{0.32\textwidth}
			\centering
			\includegraphics[scale = 0.31500]{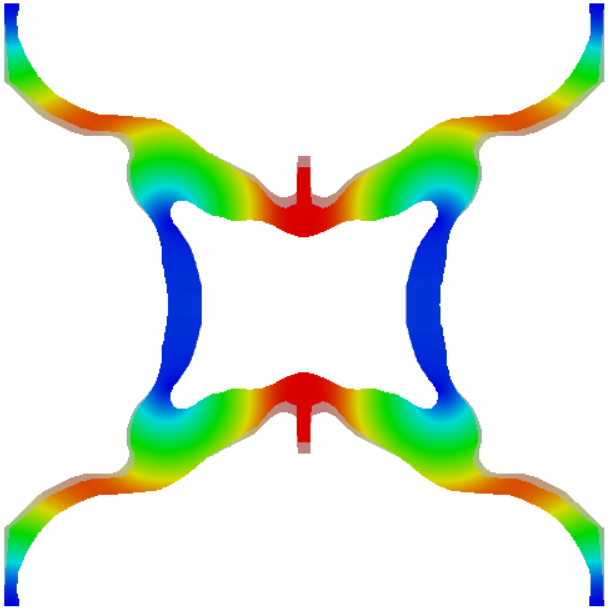}
			\caption{}
			\label{fig:CT3_10}
		\end{subfigure}
		\begin{subfigure}[t]{0.32\textwidth}
			\centering
			\includegraphics[scale = 0.350]{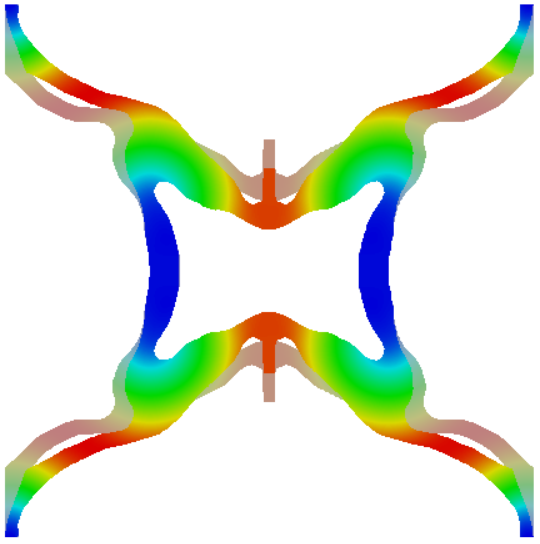}
			\caption{}
			\label{fig:CT3_25}
		\end{subfigure}
		\begin{subfigure}[t]{0.32\textwidth}
			\centering
			\includegraphics[scale = 0.350]{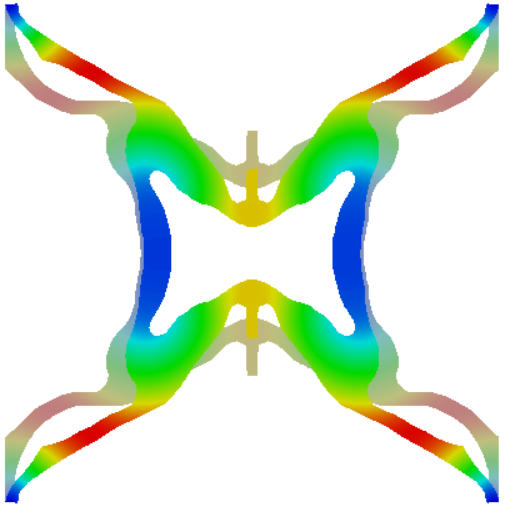}
			\caption{}
			\label{fig:CT3_50}
		\end{subfigure}
		\caption{Deformed profiles of the contractor CM at different pressure loads. CT3 Pa-CMs (\subref{fig:CT3_10}) $\SI{10}{\bar}$, (\subref{fig:CT3_25}) $\SI{25}{\bar}$, (\subref{fig:CT3_50}) $\SI{50}{\bar}$.}	\label{fig:ContractorPressure}
	\end{figure*}

	\subsection{ Extracting the optimized designs}\label{Sec:ExtOptDesigns}
	
	We present a method to extract the optimized Pa-CMs for generating CAD models and for performing further analysis with high pressure loads in ABAQUS herein. We adopt the following steps:
	\begin{enumerate}
		\item Prepare a VKT file using the optimized physical density field, nodal coordinates and element connectivity matrix.
		\item  Import the VKT file into a data visualization software, ParaView \citep{ahrens2005paraview}. Use the command  \texttt{CellDatatoPointData} which converts cell (element) information  to point (node) information.  Extract the final design  at the threshold density 0.85 value using the \texttt{IsoVolume} function, and save it as a portable network graphics (PNG) file.
		\item  Import the PNG file to a vector graphics software, InkScape. Use \texttt{Trace Bitmap} to trace the boundary and save the traced design as a DXF file.
		\item Import the  DXF file to AutoCAD software and save as an IGES file for importing into ABAQUS for the further analyses.
	\end{enumerate}
	\section{Large deformation analyses and challenges}\label{Sec:Largedeformationanalyses}
	
	In this section, the optimized compliant mechanisms are tested with high pressure loadings to investigate their behaviors under large deformation cases.  In the FE analysis, geometric nonlinearity will now be considered. In addition, instead of the linear material model, a neo-Hookean material model with the following strain energy function $W$ \citep{zienkiewicz2005finite} is employed
	\begin{equation}\label{Eq:Strainenergyfunction}
		W = \frac{G}{2}[\text{tr}\,(\bm{F}\trr{\bm{F}}) -3 -2\ln J] + \frac{\lambda}{2}(\ln J)^2,
	\end{equation}
	where $\bm{F} = \nabla_0\,\bm{u} + \bm{I}$ is the deformation gradient and $G = \frac{E_1}{2(1+\nu)}$ and $\lambda = \frac{2G\nu}{1-2\nu}$ are Lame constants.  $\nabla_0\,\bm{u}$ denotes gradient of the displacement field $\bm{u}$ with respect to reference coordinates $\bm{X}$, and $\nu$ is Poisson's ratio. $J = \det(\bm{F})$, and $\bm{I}$ is the unit tensor. Typically, rubber-like materials are used for pneumatically-actuated mechanisms \cite{schmitt2018soft} and to numerically model such materials, a neo-Hookean material description can be employed \citep{zienkiewicz2005finite}. 
	
	Using the fundamentals of the continuum mechanics, the Cauchy stress tensor $\bm{\sigma}$ can be determined from the strain energy function  noted in Eq.~\ref{Eq:Strainenergyfunction} as
	\begin{equation}\label{Eq:CauchyStress}
		\bm{\sigma} = \frac{G}{J}(\bm{F}\bm{F}^\mathrm{T}-\bm{I})+\frac{\Lambda}{J}(\ln J)\bm{I}.
	\end{equation}
	Following the nonlinear FE formulation, the displacement vector $\mathbf{u}$ is determined by solving 
	\begin{equation}\label{Eq:residualform}
		\mathbf{R}(\mathbf{u}) = \mathbf{F}_{\mathrm{int}}(\mathbf{u})-\mathbf{F}_{\mathrm{ext}}(\mathbf{u})  = \mathbf{0},
	\end{equation}
	where $\mathbf{R}(\mathbf{u})$ is the residual force and $\mathbf{F}_{\mathrm{ext}}(\mathbf{u})$ is the external force arises due to the pressure loading. The internal force vector $\mathbf{F}_{\mathrm{int}}^e(\mathbf{u})$ at the element level is determined as
	\begin{equation}\label{eq:internalforce}
		\mathbf{F}^e_\text{int} = \int_{\Omega^e}\trr{\mathbf{B}}_\text{UL}(\mathbf{u})\bm{\sigma}_e(\mathbf{u})\,d \Omega^e,
	\end{equation}
	where $\mathbf{B}_\text{UL}(\mathbf{u})$ and $\bm{\sigma}_e$  are the updated Lagrangian  strain-displacement matrix and the Cauchy stress tensor of an FE $\Omega^e$, respectively. Eq.~\ref{Eq:residualform} can be solved using a Newton-Raphson (N-R) iterative process. Note  that, $\mathbf{F}_{\mathrm{ext}}(\mathbf{u})$ varies as it arises from pressure loading which follows the surface where it is applied upon, i.e., it is a follower force and thus, contributes in the tangent stiffness of the nonlinear equations and cannot be omitted from the topology optimization. In addition, the flow coefficient matrix $\mathbf{A}^e$ (Eq.~\ref{Eq:PDEsolutionpressure}) varies with the deformation. Further analysis robustness and efficiency under distortion and/or unrealistic deformation of the low-density elements with high pressure load \citep{van2014element} pose challenges in a non-linear finite element TO setting.
	
	The elemental $\mathbf{K}_\text{ext}^e$ can be determined as (see Appendix~\ref{app:Stiffness})
	\begin{equation}\label{eq:extrnalforce}
		\mvect{K}_\mathrm{ext}^e = \int_{\Gamma_\text{p}} p\mvect{N}^\top(\mfield{n}\otimes\mfield{a}^\alpha - \mfield{a}^\alpha\otimes\mfield{n})\mvect{N}_{,\alpha} da
	\end{equation}
	where $p$ represents the magnitude of the pressure load, $\mvect{n}$ is the normal vector and $\mfield{a}^1,\,\text{and}\,\mfield{a}^2$ are the contra-variant tangent vectors to the pressure surface  $\Gamma_\text{p}$ (see Appendix~\ref{app:Surfacedescription}). $\mvect{n} =\frac{\mfield{a}_1 \times \mfield{a}_2}{||\mfield{a}_1 \times \mfield{a}_2||}$, where $\mfield{a}_1,\,\text{and}\,\mfield{a}_2$ are the co-variant tangent vectors (see Appendix~\ref{app:Surfacedescription}). $\otimes$ represents the tensor product, $da$ is the elemental area and $\alpha = {1,\,2}$. Note that with a design dependent load case, $\mathbf{n}=\mathbf{n}(\mathbf{u},\,\bar{\bm{\rho}})$ which gives additional load sensitivities and a nonlinear external force stiffness matrix. Herein, we use ABAQUS for the nonlinear finite element analyses to show the limitations of Pa-CMs optimized via linear elastic assumptions and also, to note additional challenges.
	
	The intermediate optimized designs of the inverter (IV2 and IV3), gripper (GP2 and GP3) and contractor (CT3) mechanisms  are selected for the nonlinear analyses in ABAQUS as mentioned in Sec.~\ref{Sec:Pa IV and GP}. First, the boundaries of the optimized designs are extracted, and corresponding 2D CAD models are generated (see Sec.~\ref{Sec:ExtOptDesigns}). Thereafter, using these CAD models, nonlinear FE analyses while considering follower force characteristics of the pressure loads are performed with input pressure 10 bar, 25 bar and 50 bar in ABAQUS (note, the design pressure load is 1 bar). Fig.~\ref{fig:InverterPressure}, Fig.~\ref{fig:GripperPressure}  and Fig.~\ref{fig:ContractorPressure} display the deformed profiles of the CMs with high pressure loadings. As pressure loads increase the deformation of the Pa-CMs also increase, which is expected and natural. Table~\ref{Table:T2} indicates the output displacements of the mechanisms at different pressure loads. At $\SI{50}{\bar}$, IV3 Pa-CM  experience self-contact (Fig.~\ref{fig:IV3_50}) that indicates that one may have to include \textit{self-contact conditions}  \citep{kumar2017_diss,kumar2019computational,kumar2020topology2} between the branches of the mechanisms when dealing with high pressure loadings for the large deformation cases. For the inverter mechanisms, the deformation profiles in Fig.~\ref{fig:IV3L_50} and Fig.~\ref{fig:IV3_50} are different,  and the output displacement has reduced by 90.25 \%. Likewise, the deformed continua in Fig.~\ref{fig:GP3L_50} and Fig.~\ref{fig:GP3_50} are not the same. Although the gripper still exhibits the intended functionality, the magnitude of the jaw displacement has reduced by 76.11 \%. These indicate limitations and shortcomings of the Pa-CMs obtained assuming linear elasticity concepts. Therefore, ideally, one has to include full nonlinear mechanics (with contact) within the design approach for high pressure loadings wherein Pa-CMs can experience large deformations and even self and/or mutual contact. In addition, integration of the required actuators with the mechanism/soft robot design optimization process provides an additional set of challenges \citep{cao2015toward}, which is another future research direction.
	
	\begin{table*}
		\centering
		\caption{Output displacements of the optimized  Pa-CMs under high pressure loadings}\label{Table:T2}
		\begin{tabular}{ c c c c c}
			\cline{1-4}
			\multirow{2}{*}{Mechanisms/Pressure load} & \multicolumn{1}{c}{10 bar} & \multicolumn{1}{c}{25 bar} & 50 bar &  \\ \cline{2-4}
			& \multicolumn{3}{c}{Output displacement $\Delta$ (mm)}                      &  \\ \cline{1-4}\vspace{1ex}
			IV2 Pa-CM                                 & \multicolumn{1}{c}{-31.30}  & \multicolumn{1}{c}{-39.60}  & -43.0  &  \\ \cline{1-4}\vspace{1ex}
			IV3 Pa-CM                                 & \multicolumn{1}{c}{-8.93}  & \multicolumn{1}{c}{-10.99} & -12.10  &  \\ \cline{1-4}\vspace{1ex}
			GP2 Pa-CM                                 & \multicolumn{1}{c}{-13.40}  & \multicolumn{1}{c}{-15.38}  & -15.67  &  \\ \cline{1-4}\vspace{1ex}
			GP3 Pa-CM                                 & \multicolumn{1}{c}{-7.63}  & \multicolumn{1}{c}{-12.0}  & -14.80  &  \\ \cline{1-4}\vspace{1ex}
			CT3 Pa-CM                                 & \multicolumn{1}{c}{-3.42}   & \multicolumn{1}{c}{-10.48}  & -15.30  &  \\ \cline{1-4}\vspace{1ex}
		\end{tabular}
	\end{table*}
	
\section{Closure}\label{Sec:Closure}

With the aim to bridge the gap between optimized and as-fabricated designs, this paper presents a robust density-based topology optimization approach to generate pressure-actuated  compliant mechanisms.  The robust formulation, i.e., the eroded, intermediate (blueprint) and dilated projections for the design description is employed for the first time to this problem class, in combination with a representation of the pressure loads using the Darcy law in combination with a drainage term. Well-functioning pressure-actuated robust inverter, gripper, and contractor mechanisms are obtained with different robustness levels and filter radii to illustrate different minimum length scales of the mechanisms. Single-node hinges, that cannot be fabricated but frequently appear when using the traditional topology optimization approach, are no longer found in the obtained designs. Moreover, it is observed that the robust approach leads to improved mechanism performance with better boundary crispness. The approach solves three sets of equilibrium equations for each field (pressure, displacement and virtual displacement) and uses a continuation approach for the projection parameter $\beta$ that requires a large number of the optimization iterations and thus, computational cost increases. The approach provides three physical material density vectors with one design variable vector. Intermediate designs can be used for manufacturing purposes.

The obtained optimized designs are close to binary. This eliminates the loss of performance observed when grayscale topology optimization results are post-processed into CAD geometries for fabrication. The intermediate designs are used to study behavior of the  mechanisms with high pressure loads  while accounting for geometric and material nonlinearity. It is found that the scaled linear deformation profiles and those obtained with full nonlinear analyses do not match well. Moreover, at high pressures self-contact of mechanism branches occurs. These observations indicate that for further development of topology optimization for pressure-actuated compliant mechanisms, next to the robust formulation, considering nonlinear mechanics and self-contact may be indispensable. 

\section*{Acknowledgments} 
The authors would like to thank Prof. Krister Svanberg for providing MATLAB codes of the MMA optimizer. P. Kumar acknowledges financial support from the Science \& Engineering research board, Department of Science and Technology, Government of India under the project file number RJF/2020/000023.

\section*{Declaration of Competing Interest}
The authors declare that they have no known competing financial interests or personal relationships that could have appeared to influence the work reported in this paper.

\begin{appendices}
	\numberwithin{equation}{section}
	\numberwithin{figure}{section}
	
\section{Surface description}\label{app:Surfacedescription}
To model follower forces, a ubiquitous option is to employ curvilinear coordinates setting \cite{wriggers2006computational} which is summarized herein.
\begin{figure}[h!]
	\centering
	\includegraphics[scale =1]{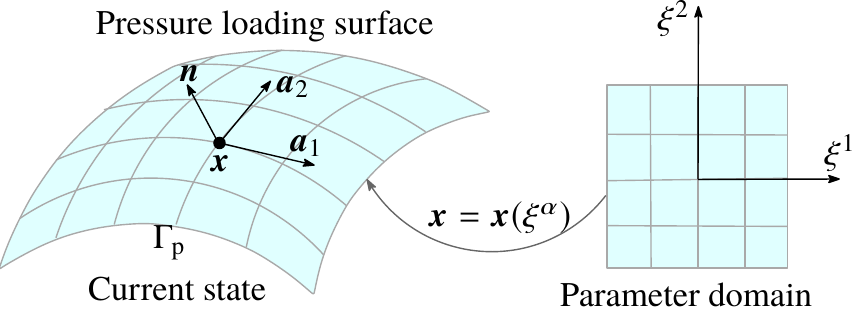}
	\caption{ A schematic diagram representing mapping}
	\label{fig:surfacedescription}
\end{figure}

Imagine	a material point $\mfield{x}\in\Gamma_\text{p}$, where $\Gamma_\text{p}$ is a pressure load surface of the evolving design\footnote{deformed configuration of $\Omega_0$} $\Omega$ (Fig.~\ref{fig:surfacedescription}). In a parametric setting, the point can be described as
\begin{equation} \label{Eq:eq1}
	\mfield{x} = \mfield{x}({{\xi}^\alpha}),\qquad {\xi}^\alpha\in \Gamma_{p}
\end{equation}
where ${\xi}^\alpha|_{\alpha = 1,2}$ are curvilinear cooridinates lying in the 2D parametric space. Please note, Greek indices in Eq.~\eqref{Eq:eq1} and henceforth adopt a set of values $\{1,\,2\}$. Repeated indices follows Einstein's summation rule. The tangent vectors at the point $\mfield{x}\in\Gamma_\text{p}$ can be determined as
\begin{equation} \label{Eq:eq2}
	\mfield{a}_\alpha= \frac{\partial \mfield{x}}{\partial \xi^\alpha},\qquad \alpha = 1,\,2
\end{equation}		
Though these tangents may not necessarily be orthonormal, they form the basis of tangent plane to $\Gamma_\text{p}$. $\mfield{a}_\alpha$ refer  co-variant tangent vectors that are related via an associated metric tensor $a_{\alpha\beta}$ as
\begin{equation} \label{Eq:eq3}
	a_{\alpha\beta}= \mfield{a}_\alpha\cdot\mfield{a}_\beta.
\end{equation} 	
The unit normal at  $\mfield{x}\in\Gamma_\text{p}$ is evaluated as
\begin{equation} \label{Eq:eq4}
	\mfield{n} = \frac{\mfield{a_}1\times \mfield{a}_2}{||\mfield{a}_1\times \mfield{a}_2||} = \frac{\mfield{a}_1\times \mfield{a}_2}{j_a},
\end{equation}
where $j_a =\sqrt{\det a_{\alpha\beta}}$ can be established \citep{wriggers2006computational,sauer2014computational}. The set $\{\mfield{a}_1,\,\mfield{a}_2,\,\mfield{n}\}$ constitutes a basis of $\mathbb{R}^3$. In addition, one can define another set of tangents $\mfield{a}^\beta$ termed contra-variant tangents as
\begin{equation} \label{Eq:eq5}
	\mfield{a}_\alpha\cdot\mfield{a}^\beta = \delta_{\alpha\beta},
\end{equation}
where $\delta_{\alpha\beta}$ is the Kronecker delta. The metric tensors of co-variant and contra-variant tangents are related as
\begin{equation} \label{Eq:eq6}
	a^{\alpha\beta} = [a_{\alpha\beta}]^{-1},
\end{equation}
where contra-variant metric tensor $a^{\alpha\beta} = \mfield{a}^\alpha\cdot\mfield{a}^\beta$. Using contra-variant tangents and the unit normal (Eq.~\ref{Eq:eq4}) one can have another set of basis vectors $\{\mfield{a}^1,\,\mfield{a}^2,\,\mfield{n}\}$ of $\mathbb{R}^3$. The co-variant and contra-variant tangents are also related as $\mfield{a}_\alpha = a_{\alpha\beta}\mfield{a}^\beta$ and $\mfield{a}^\alpha = a^{\alpha\beta}\mfield{a}_\beta$. 
\begin{figure}[h!]
	\centering
	\includegraphics[scale =1]{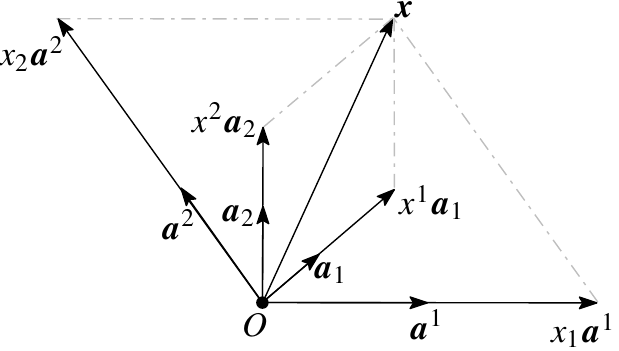}
	\caption{ Components of vector $\mfield{x}$ along co-variant and contra-variant tangents}
	\label{fig:components}
\end{figure}

Using triad $\{\mfield{a}_1,\,\mfield{a}_2,\,\mfield{n}\}$ and/or $\{\mfield{a}^1,\,\mfield{a}^2,\,\mfield{n}\}$ a vector $\mfield{x}$ can be written as (Fig. \ref{fig:components})
\begin{equation}\label{Eq:eq7}
	\mfield{x} = x^\alpha\mfield{a}_\alpha + x_n\mfield{n} = x_\alpha\mfield{a}^\alpha + x_n\mfield{n}
\end{equation}
where $x^\alpha = \mfield{x}\cdot\mfield{a}^\alpha$, $x_\alpha = \mfield{x}\cdot\mfield{a}_\alpha$ and  $x_n = \mfield{x}\cdot\mfield{n}$ represent contra-variant, co-variant and normal components of the vector, respectively. Note that  $r^\alpha = a^{\alpha\beta} r_\beta$ and $r_{\alpha} = a_{\alpha\beta} r^\beta $. Parametric derivative of the co-variant tangents can be evaluated as
\begin{equation}\label{Eq:eq8}
	\mfield{a}_{\alpha,\beta} = \frac{\partial \mfield{a}_\alpha}{\partial\xi^\beta}.
\end{equation}

\section{Evaluation of tangent stiffness matrix $\mvect{K}_\mathrm{ext}^e$} \label{app:Stiffness}

Linearization of the elemental pressure load $\mvect{F}_{\mathrm{ext}}^e$ at $\mvect{u}$ in the direction of $\Delta\mvect{u}$ is performed to evaluate $\mvect{K}_\mathrm{ext}^e$. One writes
\begin{equation} \label{Eq:app1}
	\mvect{f}_\mathrm{ext}^e (\mvect{u} + \Delta\mvect{u}) \approx \mvect{f}_\mathrm{ext}^e (\mvect{u}) + \Delta\mvect{f}_\mathrm{ext}^e (\mvect{u}),
\end{equation}
Where \citep{zienkiewicz2005finite}
\begin{equation}\label{Eq:app2}
	\mvect{f}_\mathrm{ext}^e  = \int_{\Gamma_\text{p}}\mvect{N}^\top p\mfield{n}\text{d}a,
\end{equation}
$p$ is the magnitude of the follower force. Evidently,
\begin{equation}\label{Eq:app3}
	\Delta\mvect{f}_\mathrm{ext}^e (\mvect{u}) = \int_{\Gamma_\text{p}}\mvect{N}^\top\, p\, \Delta (\mfield{n}\text{d}a) +  \int_{\Gamma_\text{p}}\mvect{N}^\top\, \Delta p\, \mfield{n}\text{d}a.
\end{equation}
Herein $\Delta p = 0$ as pressure is considered height independent. Therefore, 
\begin{equation}\label{Eq:app3.1}
	\Delta\mvect{f}_\mathrm{ext}^e (\mvect{u}) =  \int_{\Gamma_\text{p}}\mvect{N}^\top\, p\, \Delta (\mfield{n}\text{d}a).
\end{equation}
In view of Eq.~\ref{Eq:eq4}, we have
\begin{equation}\label{Eq:app3.2}
	\Delta (\mfield{n}\text{d}a)  = \Delta \left((\mfield{a_}1\times \mfield{a}_2) \text{d}\square\right) = \Delta \left(\mfield{a_}1\times \mfield{a}_2\right) \text{d}\square
\end{equation}
where $j_a\text{d}\square = \text{d}a$ and  $\text{d}\square = \text{d}\xi^1 \text{d}\xi^2$.

\textbf{B.1 Evaluation of} $\Delta(\mfield{a}_1\times\mfield{a}_2)\text{d}\square$

Herein, steps for evaluating $\Delta(\mfield{a_}1\times \mfield{a}_2) \text{d}\square$ are described.
\begin{equation}\label{Eq:app4}
	\begin{split}
		\Delta(\mfield{a_}1\times \mfield{a}_2) \text{d}\square &= (\Delta\mfield{a_}1\times \mfield{a}_2 + \mfield{a_}1\times \Delta\mfield{a}_2) \text{d}\square \\
		& = \displaystyle\sum_{I}(N_{,1}\underbrace{\Delta\mfield{u}_I\times \mfield{a}_2}_{\bm{T}_1} + \underbrace{\mfield{a}_1\times \Delta\mfield{u}_I}_{\bm{T}_2}N_{,2})\,\\
		& \qquad\qquad [\because \Delta \mfield{u}_I = \Delta\mfield{x}_I]
	\end{split}
\end{equation}
Decomposing $\Delta \mfield{u}_I$ using its components along tangential and normal direction as $\Delta \mfield{u}_I = \Delta u^1_I\mfield{a}_1 + \Delta u^2_I\mfield{a}_2 + \Delta u^n_I\mfield{n}$ allows evaluation $\bm{T}_1$ and $\bm{T}_2$ as
\begin{equation}\label{Eq:app5}
	\begin{split}
		\bm{T}_1 = \Delta\mfield{u}_I\times \mfield{a}_2 &= (\Delta u^1_I\mfield{a}_1 + \Delta u^2_I\mfield{a}_2 + \Delta u^n_I\mfield{n})\times \mfield{a}_2\big)\\
		&= j_a(\Delta u^1_I\mfield{n} -\Delta u^n_I\mfield{a}^1) \\&= j_a (\mfield{n}\otimes\mfield{a}^1 - \mfield{a}^1\otimes\mfield{n})\Delta\mfield{u}_I,
	\end{split}
\end{equation}
where $\otimes$ indicates the tensor product.
Likewise, 
\begin{equation}\label{Eq:app6}
	\bm{T}_2 = j_a (\mfield{n}\otimes\mfield{a}^2 - \mfield{a}^2	\otimes\mfield{n})\Delta\mfield{u}_I.
\end{equation}
With  $T_1$ and $T_2$, Eq.~\eqref{Eq:app4} yields
\begin{equation}\label{Eq:app7}
	\Delta(\mfield{a_}1\times \mfield{a}_2) \text{d}\square = (\mfield{n}\otimes\mfield{a}^\alpha - \mfield{a}^\alpha\otimes\mfield{n}) \mvect{N}_{,\alpha}\Delta \mvect{u}_e \text{d}a
\end{equation}
Therefore, in view of Eq.~\eqref{Eq:app3.2} and Eq.~\eqref{Eq:app7}, one writes Eq.~\eqref{Eq:app3.1} as
\begin{equation}\label{Eq:app8}
	\Delta\mvect{f}_\mathrm{ext}^e (\mvect{u}) = \mvect{K}_\mathrm{ext}^e \Delta \mvect{u}_e
\end{equation}
where
\begin{equation}\label{Eq:app9}
	\mvect{K}_\mathrm{ext}^e = \int_{\Gamma_\text{p}} p\mvect{N}^\top(\mfield{n}\otimes\mfield{a}^\alpha - \mfield{a}^\alpha\otimes\mfield{n})\mvect{N}_{,\alpha} \text{d}a.
\end{equation}
\textbf{B.2 Expression} $\mvect{K}_\mathrm{ext}^e$ \textbf{in 2D}

\begin{figure}[h!]
	\centering
	\includegraphics[scale =1]{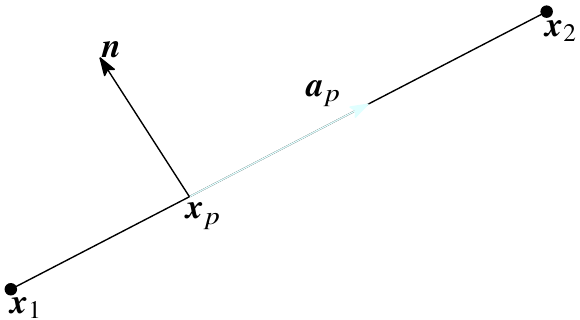}
	\caption{The edge treatment for the follower forces in 2D}
	\label{fig:aap}
\end{figure}
$\mvect{K}_\mathrm{ext}^e$ is evaluated by interpolating the edge of a quadrilateral element (Fig. \ref{fig:aap}) using linear Lagrangian shape functions (Eq.~\ref{Eq:app10}) wherein
\begin{equation}\label{Eq:app10}
	N_1 = \frac{1}{2}(1-\xi_p),\,\, N_2 = \frac{1}{2}(1+\xi_p),\,\,\xi_p\in[-1,\,1]
\end{equation}
Point $\mfield{x}_p = N_1\mfield{x}_1 + N_2\mfield{x}_2$, length of the element $l_e = ||\mfield{x}_2 - \mfield{x}_1||$,  tangent vector $\mfield{a}_p= \frac{\mfield{x}_2 - \mfield{x}_1}{l_e}$ and unit normal $\mfield{n} = \mfield{e}_3\times\mfield{a}_p$ can be evaluated. With respect of these values, we can have 
\begin{equation}\label{Eq:app11}
	\mvect{K}_\mathrm{ext}^e = \int_{\Gamma_\text{p}} p\mvect{N}^\top(\mfield{n}\otimes\mfield{a}_p - \mfield{a}_p\otimes\mfield{n})\mvect{N}_{,1} \text{d}a
\end{equation}
with $\mvect{N} = [N_1\mvect{I}_2,\,\,\,N_2\mvect{I}_2]$ and $\mvect{N}_{,1} = 0.5\times[-\mvect{I}_2,\,\,\,\mvect{I}_2]$ and $\mvect{I}_2 = \begin{bmatrix}
	1 & 0\\
	0 & 1
\end{bmatrix}$.

\end{appendices}

\end{document}